\documentclass[sigconf]{acmart}

\renewcommand\footnotetextcopyrightpermission[1]{} 
\usepackage{subcaption}
\usepackage{colortbl}
\usepackage{afterpage}

\def\BibTeX{{\rm B\kern-.05em{\sc i\kern-.025em b}\kern-.08emT\kern-.1667em\lower.7ex\hbox{E}\kern-.125emX}}
    
\copyrightyear{2019}
\acmYear{2019}
\acmDOI{TBD}
\acmISBN{TBD}

\acmSubmissionID{123-A56-BU3}

\begin{document}

%
\title{Social Cards Probably Provide For Better Understanding Of Web Archive Collections}

%
\author{Shawn M. Jones}
\email{sjone@cs.odu.edu}
\orcid{0000-0002-4372-870X}
\affiliation{%
  \institution{Old Dominion University}
  \city{Norfolk}
  \state{Virginia}
  \postcode{23529}
}

\author{Michele C. Weigle}
\email{mweigle@cs.odu.edu}
\orcid{0000-0002-2787-7166}
\affiliation{%
  \institution{Old Dominion University}
  \city{Norfolk}
  \state{Virginia}
  \postcode{23529}
}

\author{Michael L. Nelson}
\email{mln@cs.odu.edu}
\orcid{0000-0003-3749-8116}
\affiliation{%
  \institution{Old Dominion University}
  \city{Norfolk}
  \state{Virginia}
  \postcode{23529}
}

%

%
\begin{abstract}
Used by a variety of researchers, web archive collections have become invaluable sources of evidence. If a researcher is presented with a web archive collection that they did not create, how do they know what is inside so that they can use it for their own research? Search engine results and social media links are represented as surrogates, small easily digestible summaries of the underlying page. Search engines and social media have a different focus, and hence produce different surrogates than web archives. Search engine surrogates help a user answer the question ``Will this link meet my information need?'' Social media surrogates help a user decide ``Should I click on this?'' Our use case is subtly different. We hypothesize that groups of surrogates together are useful for summarizing a collection. We want to help users answer the question of ``What does the underlying collection contain?'' But which surrogate should we use? With Mechanical Turk participants, we evaluate six different surrogate types against each other. We find that the type of surrogate does not influence the time to complete the task we presented the participants. Of particular interest are social cards, surrogates typically found on social media, and browser thumbnails, screen captures of web pages rendered in a browser. At  $p=0.0569$, and $p=0.0770$, respectively, we find that social cards and social cards paired side-by-side with browser thumbnails \textit{probably} provide better collection understanding than the surrogates currently used by the popular Archive-It web archiving platform. We measure user interactions with each surrogate and find that users interact with social cards less than other types. The results of this study have implications for our web archive summarization work, live web curation platforms, social media, and more.

\end{abstract}

%
%


%

%

%
\maketitle

\begin{figure}[t]
	\centering
	\includegraphics[width=0.5\textwidth]{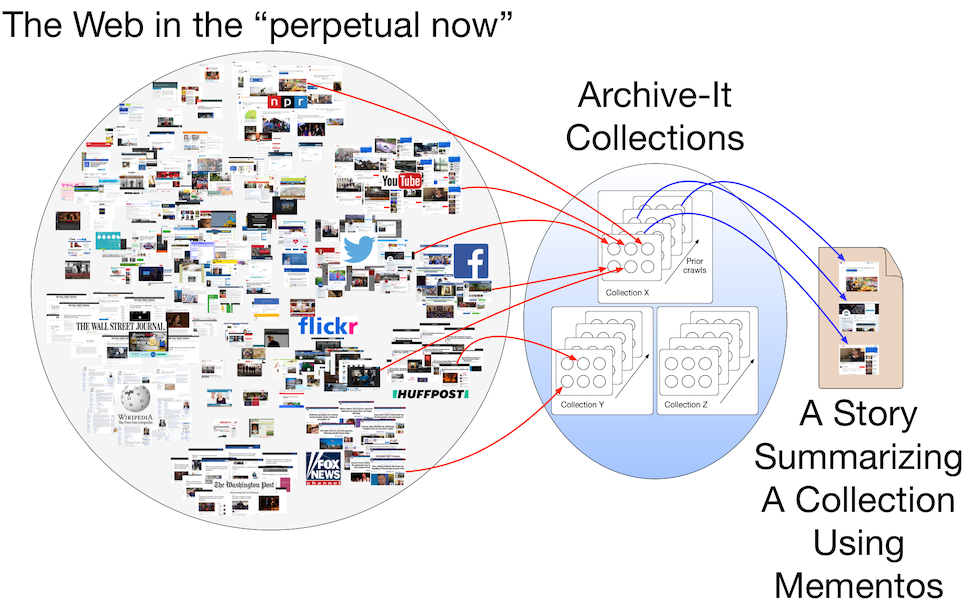}
	\caption{Web archive collections provide meaning by sampling specific resources from the web based on a theme. These collections are still large, and their mementos are observations from specific points in time. We want to sample mementos these collections to produce a much smaller story using surrogates, but which surrogate works best for a story?}
	\label{fig:DSA-overview}
\end{figure}

\begin{figure*}
    \centering
    \begin{subfigure}[t]{0.5\textwidth}
        \includegraphics[width=\textwidth]{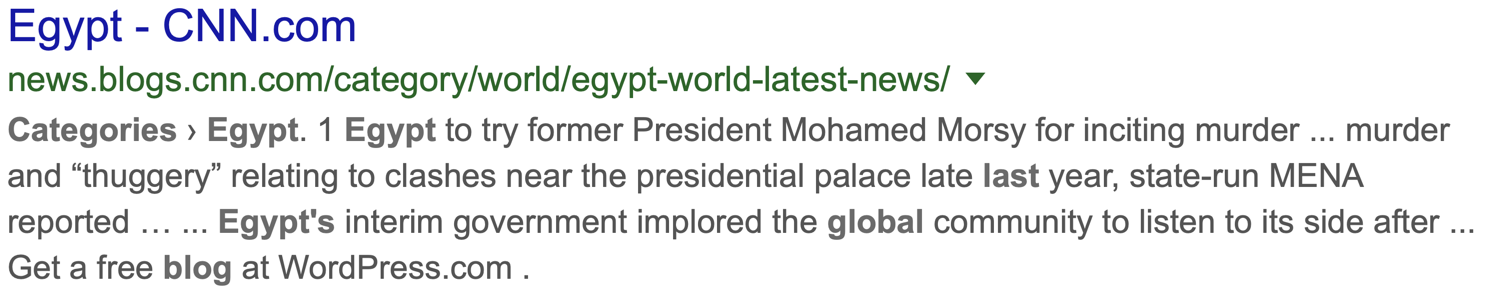}
        \caption{A Google search result surrogate}
        \label{fig:google_surrogate}
    \end{subfigure}
    \qquad
    \begin{subfigure}[t]{0.3\textwidth}
		\includegraphics[width=\textwidth]{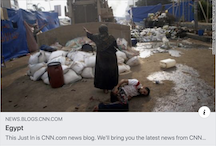}
        \caption{An example of the same URI rendered in a Facebook social card}
        \label{fig:facebook_surrogate}
    \end{subfigure}
	\caption{Two surrogates from different services for the URI \url{http://news.blogs.cnn.com/category/world/egypt-world-latest-news/}. Both contain the URI domain name and text snippet from that web pages. Social cards also contain a striking image selected from that page.  Google search results often contain the full URI.}
	\label{fig:two_surrogates}
\end{figure*}

\begin{figure*}[t]
	\centering
	\includegraphics[width=\textwidth]{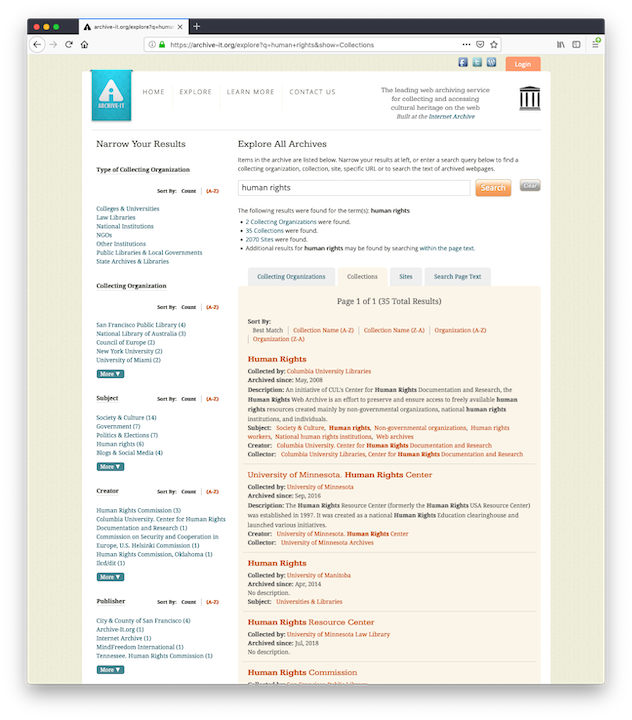}
	\caption{The search term ``human rights'' returns 35 Archive-It collections}
	\label{fig:human_rights_search}
\end{figure*}

\begin{figure*}[t]
	\centering
	\includegraphics[width=\textwidth]{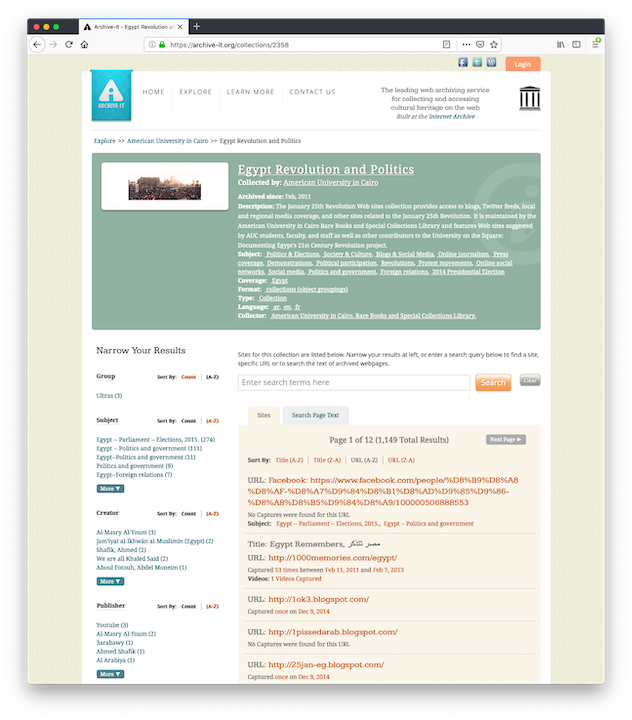}
	\caption{The Archive-It interface for collection \emph{Egypt Revolution and Politics} (ID 2358)}
	\label{fig:archiveit_page_example}
\end{figure*}

\section{Introduction}
Curators create web archive collections to preserve pages, thereby preventing link rot or content drift, according to a particular theme or collection development policy. Such collections have been used by historians \cite{milligan_book_2019}, journalists \cite{hafner2017}, and other researchers \cite{MEET:MEET14504801096} to understand the details of particular events, subjects, or even the changes in an organization. These collections are often built using tools like the Internet Archive's subscription-based service Archive-It\footnote{\url{https://archive-it.org}}. When such collections are encountered by those who did not build them, how are these third-parties to know what they contain?

Web pages exist in the ``perpetual now'' and are updated with new content as needed. The Memento protocol \cite{rfc7089} uses the term \textbf{original resource} to refer to the current version of the web page on the live web. Curators build web archive collections by employing software known as a \textbf{crawler}. A crawler visits the original resource, and the representation captured at crawl time is known as a \textbf{memento}, a version of the page now in the archive that will no longer change even if the live web version changes. Curators create a collection by choosing seeds, URLs of original resources from which to begin the crawl. Depending on the crawl parameters, the collection could include additional original resources that are not seeds (e.g., pages linked from a seed). Figure \ref{fig:DSA-overview} displays a simplified view of this collection building. Curators select seeds based on a theme. They crawl these seeds at different points in time, thus each seed produces multiple mementos of that seed, with each memento representing the seed page at a different point in time. In addition, a curator can instruct the software to follow all links from each page, resulting in many more mementos linked from the seed and then linked from those pages. For example, if a seed has three links to pages with three links each, a single crawl can lead to 13 documents being added to the collection. If this same seed is crawled three times, then 39 documents are added to the collection. This process causes web archive collections to grow to hundreds or thousands of documents.


Inspired by the work of AlNoamany et al. \cite{alnoamany2017}, we want to provide users with a visualization that allows them to understand a collection so that they can determine if the time spent evaluating these thousands of documents is worthwhile. Rather than synthesizing additional material, we want to intelligently sample $k$ mementos from the $N$ mementos that are in the collection, such that $k \ll N$. Our $k$ mementos become a \textbf{story} summarizing the collection. The right side of Figure \ref{fig:DSA-overview} displays the storytelling part of the process. AlNoamany's work visualized mementos using the now-defunct social media service Storify \cite{jones2017storify, storify_eol}, but was this the best interface? Given a sample of $k$ mementos, how do we effectively visualize these stories so that a user understands the underlying collection?

Existing information retrieval (IR) research has focused on the concept of providing each search result to a user as a \textbf{surrogate} of the underlying web page. Figure \ref{fig:google_surrogate} displays a surrogate from a Google search engine result page. Surrogates are used by search engines to answer a user's question of ``Will this link meet my information need?'' Social media uses surrogates as well. Figure \ref{fig:facebook_surrogate} displays the same URI rendered as a Facebook social card. In social media, surrogates answer the question of ``Should I click on this?'' The differences in use cases are subtle. Each surrogate is a summary of the page, often providing images, text, and metadata. We wish to use surrogates as well, but our use case is different from search engines and social media. In social media, the user focuses on a single surrogate. In IR, they compare many surrogates to each other, but for discriminating between documents. We want to provide them with a cohesive story using the combination of many surrogates together as a single unit. Using a visualization of not one, but $k$ surrogates, we want to answer the user's question of ``What does the underlying collection contain?'' The mementos in this visualization are not search results, but a product of this automatic sampling. We wish to challenge conventional wisdom beyond aesthetics. Our goal is to demonstrate the utility of a given surrogate for our web archive collection use case. There are many types of surrogates. Which one best conveys the concepts of the underlying collection?

In this pilot work, we consider six different types of surrogates and how well they might work to convey understanding of a collection. We compare the existing Archive-It surrogates, thumbnails of page screenshots, social cards, and three combinations of social cards and thumbnails. Our hypothesis is that surrogates with more information drawn from the source document produce better results, both in terms of time and understanding. Because we are evaluating surrogates for use in collection understanding rather than search engine result performance, we consider this to be a unique contribution. Overall, our results show that the type of surrogate does not influence the time to complete the task, but social cards ($p=0.0569$) and social cards side-by-side with thumbnails ($p=0.0770$) \emph{probably} provide better collection understanding than the existing text-based Archive-It interface(Figure \ref{fig:archiveit_page_example}). We find that our participants interact most with the social card side-by-side with thumbnails and second most with screenshots alone. While Archive-It is our focus, our results can be applied to other web archiving platforms, such as Webrecorder\footnote{\url{https://webrecorder.io/}}. These results are important in understanding not only which surrogate performs best for our web archiving summaries, but also for social media, live web curation platforms, and bookmarking applications as well.

\begin{figure}[t]
	\centering
	\includegraphics[width=0.5\textwidth]{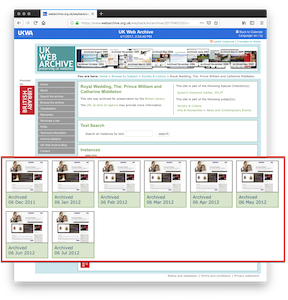}
	\caption[The UK Web Archive uses thumbnails to display the content of mementos for a single resource]{The UK Web Archive uses thumbnails to display the content of mementos for a single resource\footnotemark (thumbnails have been magnified and boxed in red for emphasis).}
	\label{fig:ukwa_thumbnails}
\end{figure}

\begin{figure*}
	\centering
	\begin{subfigure}[t]{\textwidth}
		\centering
    	\includegraphics[height=3in]{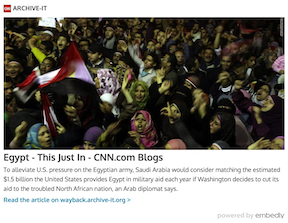}
    	\caption{A social card produced by Embed.ly for a memento from the \emph{Egypt Revolution and Politics} collection. Did CNN or Archive-It create this content? Should the reader be suspicious?}
    	\label{fig:confusing_embedly}
	\end{subfigure}

	\bigskip
	
	\begin{subfigure}[t]{\textwidth}
		\centering
    	\includegraphics[width=\textwidth]{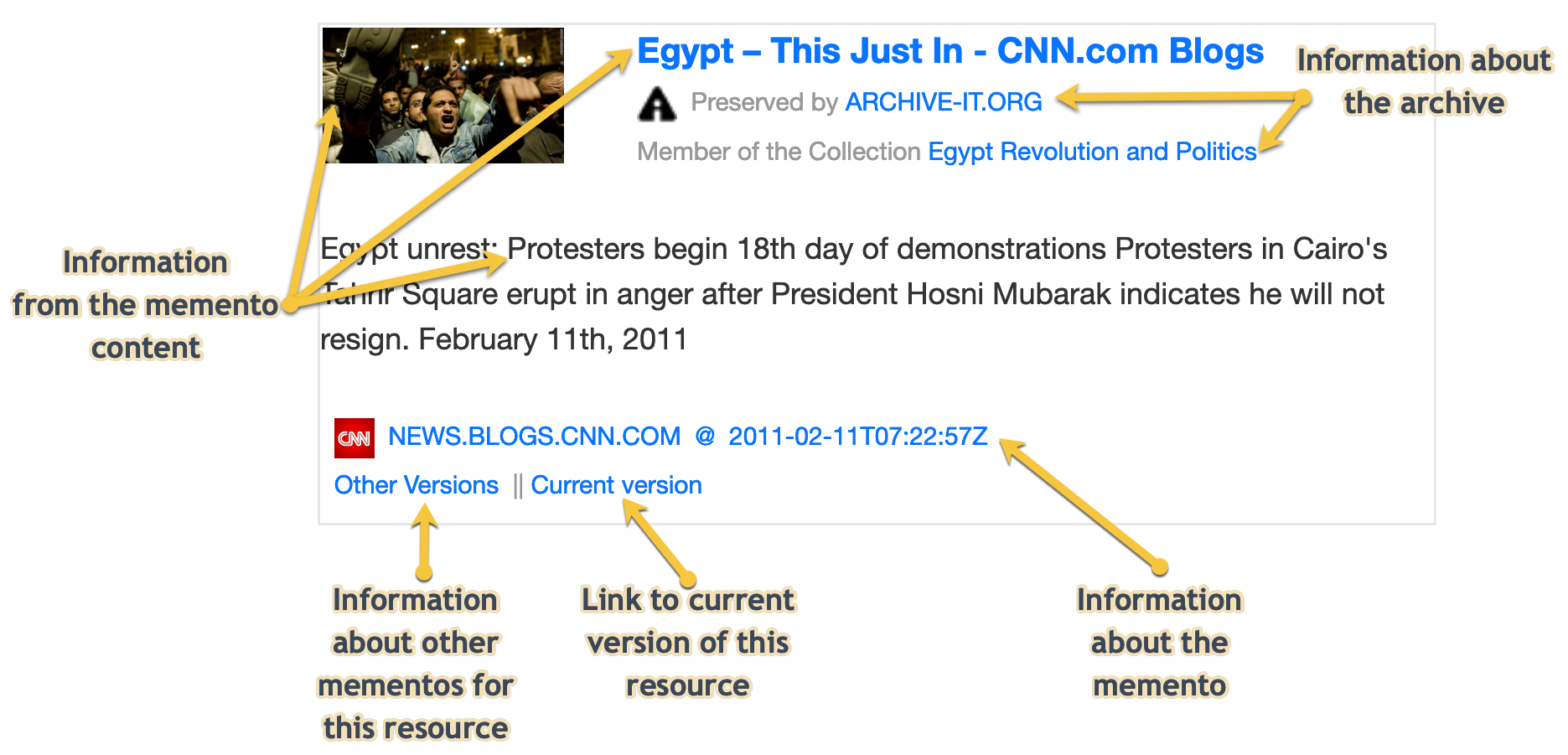}
    	\caption{We developed MementoEmbed to create surrogates that address such uncertainty. Note how the archive information is separate from the original resource information, thus avoiding confusion.}
    	\label{fig:mementoembed_annotated}
	\end{subfigure}
\caption{The same URI-M from the \emph{Egypt Revolution and Politics} collection, visualized in Embed.ly, and then visualized with MementoEmbed and annotated.}
\label{fig:mementoembed_to_the_rescue}
\end{figure*}

\section{Background}

With more than 8,000 collections \cite{Jones_Nelson_Weigle_2018_2} by the end of 2017, the Internet Archive's Archive-It is the largest web archive collection platform. It allows curators to easily select seeds and control crawling behavior. By default, Archive-It starts each crawl at a seed and creates mementos of other linked documents from the same web site until it reaches a preconfigured document count, storage limit, or time limit. With each curator's subscription comes a pre-established data storage limit, bounding the size of all of their collections. Thus, it is in their best interest not to create an excessive number of mementos. Curators can change crawling behavior in a variety of ways ranging from asking Archive-It to only crawl the single page to expanding the crawling scope to include connected web sites. It is difficult for an outsider to know which crawling behavior was selected at the time of crawl without again crawling the resulting mementos.

Archive-It provides a search interface allowing a user to find collections matching certain key words. Figure \ref{fig:human_rights_search} demonstrates that searches for topics such as ``human rights'' return more than 30 collections. When a third-party user accesses one of these collections, they are greeted by an interface like that shown in Figure \ref{fig:archiveit_page_example}. This interface is seed-centric, driving users to explore the collection first via the URLs of seeds and the metadata supplied by curators. To understand the collection, a user must follow a link from this seed interface to a list of mementos for that seed. These mementos are accessible via URIs like any other web resource. To differentiate them from original resource URIs, we refer to memento URIs with the Memento protocol \cite{rfc7089} standard nomenclature \textbf{URI-M}. The user clicks on a link to a URI-M from that list to then read its contents. The user can then follow links to other mementos until they reach a page that was not archived. From there, they can select another memento from the same seed or start again with a link from the seed. This is a tedious process, requiring going through thousands of documents to understand the collection. If a human is trying to decide between many collections, they would need to go through many documents one-at-a-time using this interface. To narrow down the number of mementos to review, a user can employ the Archive-It search engine on a single collection, but they must know enough about the collection to form a query.

Each Archive-It collection has a page, shown in Figure \ref{fig:archiveit_page_example}, that allows end users to view metadata about the collection and search within its contents via traditional IR techniques such as facets and search forms. Metadata is optional and may not be present on seeds or even entire collections.

\footnotetext{Example URI: \url{https://www.webarchive.org.uk/wayback/en/archive/20170401203440/http://www.webarchive.org.uk/ukwa/target/63275058/source/subject}}

Each seed (not memento) has its own surrogate in the Archive-It interface. Curators can enhance these surrogates with metadata, but again it is optional. The storage of this metadata is already handled by the database backend of Archive-It.  We analyze Archive-It surrogates and discuss their metadata in Section 4.

Browser thumbnails are screen captures of a web page rendered in a browser. Kopetzky demonstrated the use of thumbnails as surrogates as early as 1999 \cite{kopetzky_visual_1999}. Shown in Figure \ref{fig:ukwa_thumbnails}, the UK Web Archive uses browser thumbnails as surrogates for mementos in its collections. These browser thumbnails are also used by other collection visualization tools, such as TMVis \cite{weigle2017} and What Did It Look Like? \cite{nwala2015}, to show how seeds change over time. Even though their generation can be automated with tools like Puppeteer\footnote{\url{https://pptr.dev}}, browser thumbnails require significant resources to create.  Generation involves launching a browser, loading the page, including all images and scripts, and then taking a screenshot of that page. In addition to the memory and processing needed, thumbnails also require multiple network connections to retrieve all resources for a page. In aggregate, browser thumbnails can also be costly to store, leading the UK Web Archive to only store thumbnails for seeds, but not linked pages \cite{jackson_discussion_2019}. This cost in time and resources has led to research that focuses on optimizing the selection of mementos worthy of thumbnails \cite{AlSum2014}. The UK Web Archive uses thumbnails only 98 pixels wide. Because we seek to evaluate understanding, the thumbnails in this study are 208 pixels wide, established as the optimal size for high recognition by Kaasten \cite{kaasten2002}.

A common surrogate found in social media is the social card, like the Facebook example in Figure \ref{fig:facebook_surrogate}. Social cards typically contain an image selected from the underlying web page, the title of that page, and some text sampled from the page. Social cards can require fewer HTTP requests than thumbnails. They extract existing content from the page and do not require the time and space required to create and store new content, such as a thumbnail. The popularity of social cards has encouraged both Twitter and Facebook to recommend specific HTML metadata fields so that authors can control how cards are generated from their pages\footnote{\url{http://ogp.me/}}\textsuperscript{,}\footnote{\url{https://developer.twitter.com/en/docs/tweets/optimize-with-cards/guides/getting-started}}. We know of no web archives that currently use social cards as surrogates for their mementos.

Services such as Embed.ly\footnote{\url{https://embed.ly}} exist to produce social cards of live web resources. When used to generate social cards for mementos, they create a poor or confusing experience for users (Figure \ref{fig:confusing_embedly}). For this reason, we have developed MementoEmbed, an archive-aware platform that accepts a URI-M and then generates either a social card or a thumbnail for that memento \cite{jones2018mementoembed}. In addition to the image, title, and text provided by most social cards, MementoEmbed also provides the date and time of the observation leading to the memento, its original domain name and favicon, the name and favicon of the web archive holding it, and links to other versions of this same page (Figure \ref{fig:mementoembed_annotated}). Most of this data comes from the underlying Memento protocol supported by many web archives \cite{rfc7089}. MementoEmbed is used to generate the social cards and thumbnails used in our study.

To evaluate our surrogates in terms of understanding, we have several requirements. We must recruit participants and we must also provide a consistent environment to evaluate their understanding. We must then evaluate how well the participants demonstrate that they understand some elements of the underlying collection by viewing the surrogates. 

To recruit a sufficient number of participants for this study, we turned to Mechanical Turk (MT). MT provides a web interface for participants to view information and fill out surveys. MT participants are paid for their submissions. Each assignment in MT is referred to as a Human Interface Task (HIT). MT has been used in other visualization studies with great success. It has allowed researchers to verify earlier results with a larger set of participants \cite{kosara_mechanical_2010, 10.1371/journal.pone.0121595}, and others have used it to test the effectiveness of new visualization techniques \cite{heer2010}. As our surrogates are visualizations of underlying mementos, this past support provides confidence in MT as a recruitment tool.


Bloom's taxonomy \cite{Bloom1956} and Anderson and Krathwohl's later revision \cite{Anderson2001} provide definitions for different levels of cognitive effort with respect to learning a subject. Kelly applies these concepts to the development of search tasks \cite{kelly2015} for IR studies. In our study, we focus on two levels from this taxonomy. The \emph{remember} process requires that the participant demonstrate the ability to identify and retrieve specific facts. The \emph{understanding} process requires that the participant infer and construct additional meaning based on what they have learned already. We evaluate a user's ability to remember by giving them 30 seconds to view a visualization before presenting them with a question. We evaluate their ability to understand by asking them to select which mementos from a list likely come from the collection that they just viewed.

\section{Related Work}

Summarization of Archive-It collections using surrogates was pioneered by AlNoamany et al. \cite{alnoamany2017}. She focused on the use of Storify as the target visualization platform, but Storify has been shut down \cite{jones2017storify, storify_eol}. Storify used social cards exclusively, and AlNoamany et al. did not evaluate other surrogate types.

All of the following studies evaluated surrogates in terms of search engine result relevance. In 2001, Woodruff et al. attempted to improve upon the browser thumbnail by introducing the ``enhanced thumbnail'', which also included highlighted and enlarged text to further convey aboutness \cite{woodruff2001}. As search result surrogates, she discovered that thumbnails outperformed text alone, and enhanced thumbnails outperformed thumbnails. Unfortunately, the discovery and enlargement of text made enhanced thumbnails computationally expensive to create. In 2009, Teevan et al. further sought to replace the thumbnail with the ``visual snippet'' \cite{teevan2009}. Visual snippets consist of a 120-by-120 pixel image representing the page  constructed from an internal image, the title, and a logo. Her user testing showed that they performed better than thumbnails. She also demonstrated that text alone performed better than thumbnails. Dziadosz and Chandrasekar \cite{dziadosz2002} found that text alone combined with thumbnails performed better than merely text alone and that text alone performed better than thumbnails alone. Aula et al. \cite{aula2010} discovered no difference in performance between text alone and thumbnails. She also examined text combined with thumbnails and found no difference in performance.

In more recent years, social cards have become a topic of study. Al Maqbali et al. \cite{almaqbali2010} discovered no performance difference between text combined with thumbnail, social card, or text alone. Loumakis discovered no performance difference between text snippets or social cards \cite{loumakis2011}. Capra et al. \cite{capra2013} discovered that social cards were barely more performant than text snippets for search.

These studies all consider how well these surrogates perform for the purpose of relevance judgements in search results. The surrogate only needed to answer a single question for the user: ``Will this link meet my information need?'' We differ by considering how well the surrogates themselves convey understanding when presented together as a story summarizing a web archive collection, answering the question of ``what does the underlying collection contain?'' Our study also provides a unique contribution in this space, as none of these prior studies compare browser thumbnails to social cards directly.

We have chosen MT as a recruitment tool for evaluating our visualizations. Kittur et al. \cite{kittur_can_2008} evaluated using MT for complex tasks. He cautions that participants are encouraged to complete tasks quickly to increase their rate of pay, but sometimes this results in nonsense answers, thus it is ``best suited for tasks in which there is a bonafide answer''. His study showed that MT could be used for the complex task of rating the quality of Wikipedia articles, producing similar results to human Wikipedia curators. Bartneck et al. \cite{10.1371/journal.pone.0121595} asked participants to rate expressions on LEGO minifigure faces, and discovered that MT participants performed as well as participants from in-person studies. Heer et al. \cite{heer2010} repeated a well known visualization study using MT participants. Heer showed participants different visualizations, and asked them to identify the smaller of two marked values. Heer then asked them to estimate what percentage the smaller was of the larger. Heer's results were consistent with the original study showing that position outperformed length in terms of human cognition. After establishing that MT participants were consistent with in-person studies, Heer went on to evaluate new visualizations. Micallef et al. \cite{micallef2012} evaluated different visualization techniques for understanding the results of Bayesian problems. She confirmed that MT participants did not perform better with any of the visualizations.  Yu et al. \cite{yu2013} used MT participants to discover which pictograms better described at-home medical procedures. The success of these studies informs our choice of MT as a recruitment tool.

\begin{figure}[t]
\includegraphics[width=0.5\textwidth]{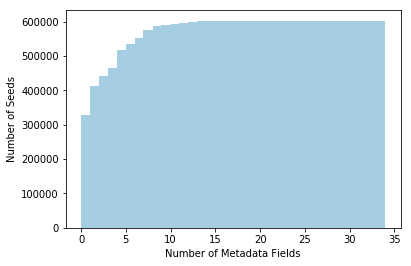}
\caption{A cumulative histogram of the number of metadata fields in user per Archive-It surrogate.}
\label{fig:seeds_metadata_hist}
\end{figure}

\begin{figure}[t]
\includegraphics[width=0.5\textwidth]{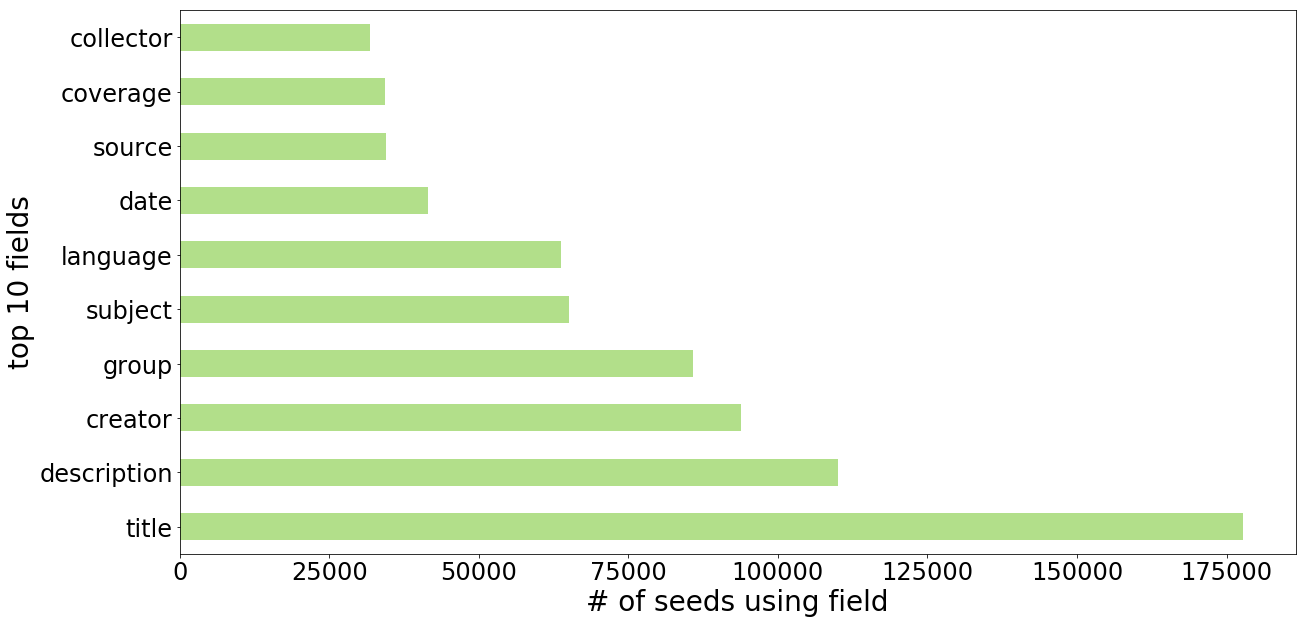}
\caption{The top 10 metadata fields used by seeds in Archive-It.}
\label{fig:top10fields}
\end{figure}

\begin{figure}[t]
\includegraphics[width=0.5\textwidth]{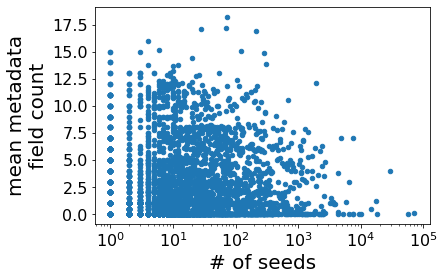}
\caption{A plot of mean metadata field count per collection vs. the number of seeds in an Archive-It collection for 5,857 public Archive-It collections.}
\label{fig:mean_seed_metadata_fields_vs_seedcount}
\end{figure}

\begin{figure}[t]
\includegraphics[width=0.5\textwidth]{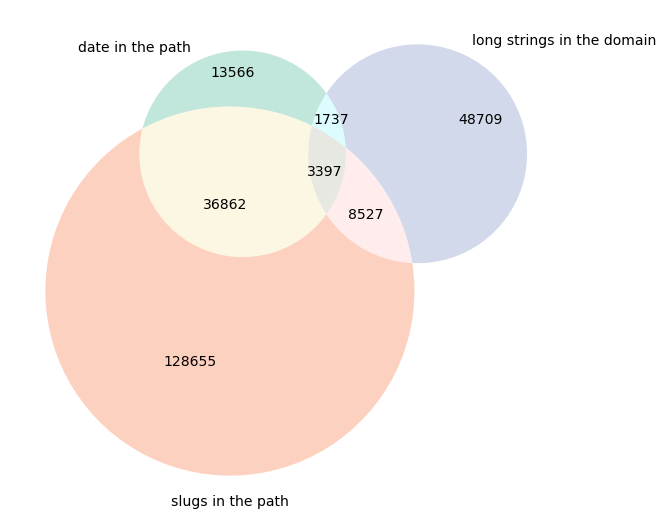}
\caption{Overlap of different information classes of Archive-It seed URIs.}
\label{fig:archiveit_surrogate_euler}
\end{figure}

\begin{figure}[t]
\includegraphics[width=0.5\textwidth]{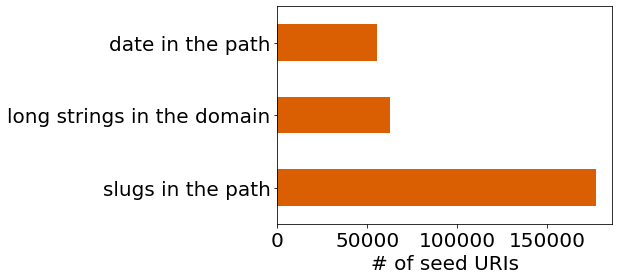}
\caption{A bar chart quantifying the potential information detected in Archive-It seed URIs.}
\label{fig:slug_information}
\end{figure}

\section{Evaluation of Archive-It Surrogates}

Before evaluating discussing the results of evaluating different surrogates against each other, we first quantify the information available from Archive-It surrogates. Rather than accepting colloquial reports about the variation in Archive-It surrogates, we used the aiu Python package \cite{jones2018aiu} to collect the metadata of 5,857 public Archive-It collections in March 2019. Our goal was to understand the amount of metadata available with most Archive-It surrogates.

Curators may also supply metadata for these seeds, but not their mementos \cite{praetzellis_maria_2016}. The metadata available is based on structural vocabulary provided by Archive-It. Most of these fields come from Dublin Core \cite{dublincore}, with some Archive-It specific fields like \texttt{group}. The curator can also supply fields from their own freeform vocabulary. Archive-It surrogates contain two sources of metadata: the archiving process and the original resource. A seed's \textbf{minimal Archive-It surrogate} contains information from the archiving process: the seed's URL, the dates of the first and last memento, and the number of mementos available. The third surrogate in Figure \ref{fig:archiveit_page_example} is an example of a minimal surrogate. The \texttt{title} is an example of metadata derived from the original resource. It may be manually extracted at the time the seed is added to the collection or may be manually added later by the curator. These original resource fields are optional.

In addition to metadata being nonexistent, it is also inconsistently applied among surrogates, as seen in the Figure \ref{fig:archiveit_page_example}. As shown in Figure \ref{fig:seeds_metadata_hist}, from the 602,944 seeds gathered, 329,178 have no metadata, meaning that 54.60\% of seeds are represented by the minimal Archive-It surrogate. These seeds convey only the URL and information from the archiving process. Of the 84,558/602,944 (14.02\%) seeds using one metadata field, the top three fields in use are \texttt{group} (40,144/84,558), \texttt{title} (29,175/84,558), and \texttt{coverage} (8,665/84,558).  \texttt{Group} allows the user to create sub-collections of seeds.  The \texttt{coverage} field corresponds to the Dublin Core field of the same name.


Figure \ref{fig:top10fields} shows the top ten fields in use by any seed, regardless of the number of fields per seed. In this case \texttt{title} is the most widely used metadata field, being present in 177,680/602,944 (29.5\%) seeds. The \texttt{description} field is in use by 110,065/602,944 (18.3\%) seeds. If metadata fields are provided by the curator, such as a \texttt{title} or \texttt{description}, the Archive-It surrogate begins to resemble surrogates typically found in search engine results. The two fields together are used on 75,575/602,944 seeds, meaning that 12.53\% of Archive-It seeds contain the same metadata fields as a Google surrogate.

Some collections, such as \emph{Government of Canada Publications} (ID 3572)\footnote{\url{https://archive-it.org/collections/3572}}, have hundreds of thousands of seeds, making the addition of metadata a costly proposition in terms of manual time and effort. Does this cost affect the behavior of the curator? For each collection, we counted how many metadata fields were applied to all seeds in the collection, regardless of size. We then divided the number of fields counted by the number of seeds to produce the mean metadata field count per collection. Figure \ref{fig:mean_seed_metadata_fields_vs_seedcount} shows a point for each collection where the y-axis is the mean metadata field count and the x-axis is the number of seeds in $\log_{10}$ scale. This graph displays a pattern whereby an increase in the number of seeds corresponds to a decrease in the number of metadata fields used to describe those seeds. This matches our intuition that because each metadata field requires some level of effort to maintain, the curator supplies fewer metadata fields as the number of seeds increases. The mean metadata field count for 3,096/5,867 (52.86\%) collections is $0$, again indicating that a majority of collections only contain minimal Archive-It surrogates.

These results appear to support our intuition that many of the Archive-It surrogates contain little information, but do they? How much information can be gathered from the seed URIs? As noted, there are many collections about the same topic, so there is some overlap in choice of seed URIs by different curators. There are 14,179 repeated seed URIs across Archive-It collections, meaning that only 588,749 unique seed URIs exist in Archive-It. From those seed URIs, we employed regular expressions from Alkwai's work \cite{alkwai_dissertation_2019} to detect different forms of crude information available in the seed URIs from Archive-It. As shown in Figure \ref{fig:archiveit_surrogate_euler}, it is possible for a seed URI to belong to all three information classes.

Our regular expressions only detected dates in the paths of 55,924/588,749 (9.44\%) seed URIs. Such dates are typically the publication dates of blog posts or news articles. Dates can provide the viewer with a concept of aboutness with respect to the time period of a collection.

Long strings may indicate the presence of phrases or sentences. Long strings are defined as any string greater than five alphabetic characters followed by an underscore or other separator, followed by another set of five alphabetic characters. We discovered 62,370/588,749 (10.59\%) URIs contained long strings in their domain names.

We borrow the term \textbf{slug} from journalism, where it indicates a shortened title for an article. Slugs are detected in the path part of a URI using the same rules as long strings. We discovered that 177,441/588,749 (30.14\%) seed URIs contained slugs in their path.

Figure \ref{fig:slug_information} displays the results of this analysis. These results indicate that, in spite of missing metadata, information can still be gleaned from the URIs found in Archive-It surrogates.

\section{Comparing Surrogates}

\subsection{Methodology}

We conducted prototype studies in Fall 2018. Rather than developing reading comprehension questions or constructing artificial search tasks, we chose something more easily measurable and verifiable: a checklist of known correct or incorrect items that we could keep consistent between participants viewing the same collection. This would more directly let us compare their performance. These prototypes taught us to avoid tasks that would overly favor one surrogate over another. We also wanted to ensure that the participants did not rely on their own knowledge and instead used the information from the visualization they were presented to answer the question.

In January 2019 we presented 120 MT participants with a link to a survey hosted at Old Dominion University. We produced four stories represented by six different surrogates for 24 different combinations of surrogates and stories. This gave us five participants per story-surrogate combination, providing 20 participants per surrogate type. The MT participants were required to have the Master Turker qualification and an acceptance rate of greater than 95\%. To control for the effects of learning \cite{kelly_methods_2007}, we employed UniqueTurker\footnote{\url{http://uniqueturker.myleott.com}} to ensure that the same participant did not provide results for multiple surveys. Each participant was paid \$0.50 to complete the task.

After reading the instructions, each participant was given 30 seconds to view a story using a given surrogate. They were then asked a question about what they had just seen. As is common practice for externally hosted surveys on MT, once they submitted their results, they were given a completion code for the MT HIT so that we could map their results to those collected by our survey.


\begin{figure*}[htbp]
\includegraphics[height=5in]{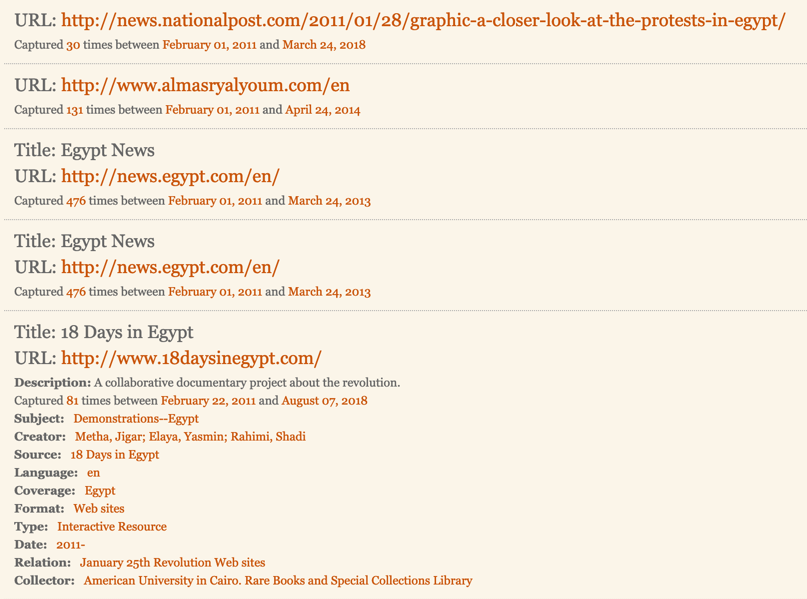}
\caption{A screenshot of part of a story constructed from a sample of $16$ mementos drawn from the $80,484$ mementos found in collection \emph{Egypt Revolution and Politics} (ID 2358), visualized as surrogates from the Archive-It Interface.}
\label{fig:example_story_archiveitlike}
\end{figure*}

\begin{figure*}[htbp]
\includegraphics[height=4in]{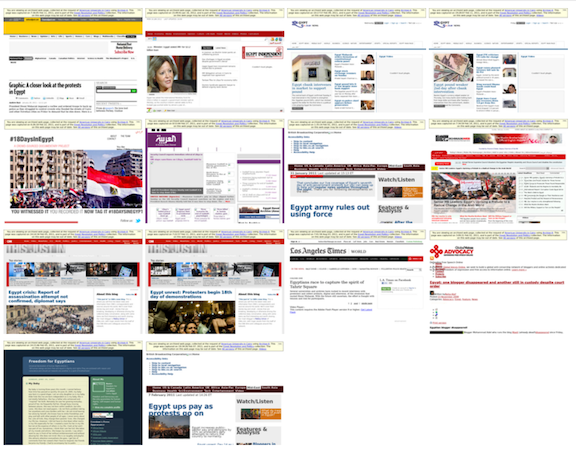}
\caption{A screenshot of a story constructed from a sample of $16$ mementos drawn from the $80,484$ mementos found in collection \emph{Egypt Revolution and Politics} (ID 2358), visualized using browser thumbnail surrogates.}
\label{fig:example_story_thumbnails}
\end{figure*}

\begin{figure*}[htbp]
\includegraphics[height=6in]{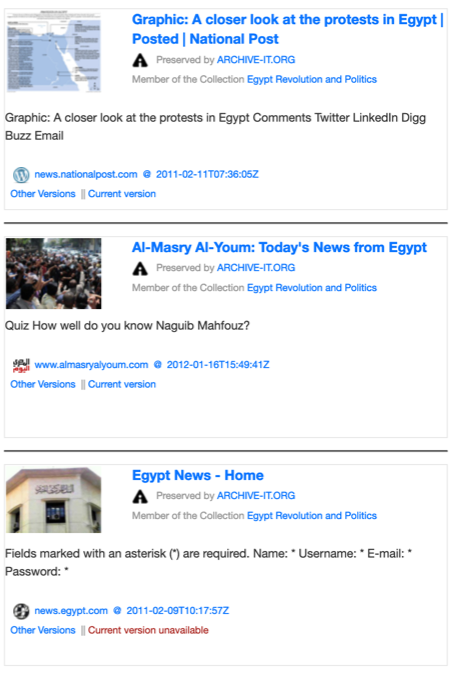}
\caption{A screenshot of part of a story constructed from a sample of $16$ mementos drawn from the $80,484$ mementos found in collection \emph{Egypt Revolution and Politics} (ID 2358), visualized using social card surrogates.}
\label{fig:example_story_socialcard}
\end{figure*}

\begin{figure*}[htbp]
\includegraphics[height=6in]{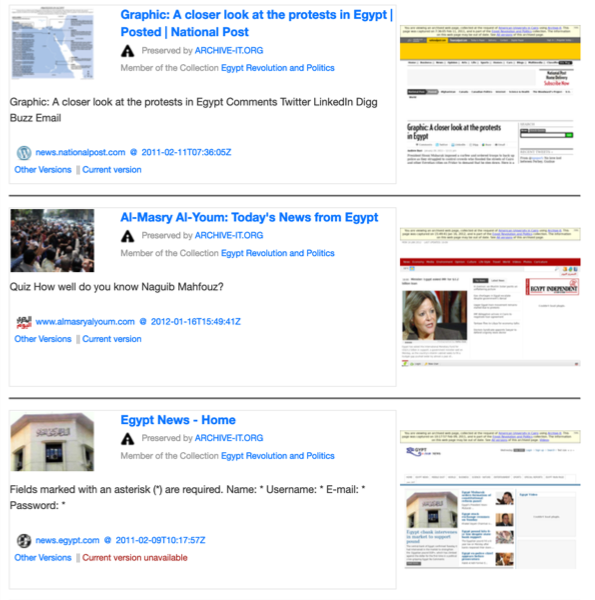}
\caption{A screenshot of part of a story constructed from a sample of $16$ mementos drawn from the $80,484$ mementos found in collection \emph{Egypt Revolution and Politics} (ID 2358), visualized using sc+t surrogates.}
\label{fig:example_story_socialcardthumbnail}
\end{figure*}

\begin{figure*}[htbp]
\includegraphics[height=6in]{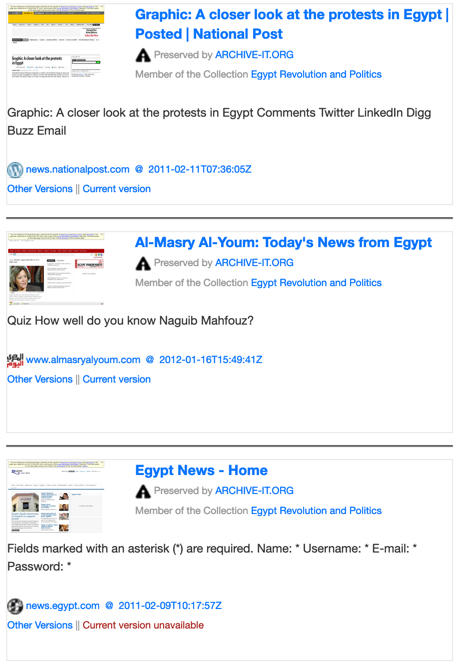}
\caption{A screenshot of part of a story constructed from a sample of $16$ mementos drawn from the $80,484$ mementos found in collection \emph{Egypt Revolution and Politics} (ID 2358), visualized using sc/t surrogates.}
\label{fig:example_story_socialcardwiththumbasimg}
\end{figure*}

\begin{figure*}[htbp]
\includegraphics[height=6in]{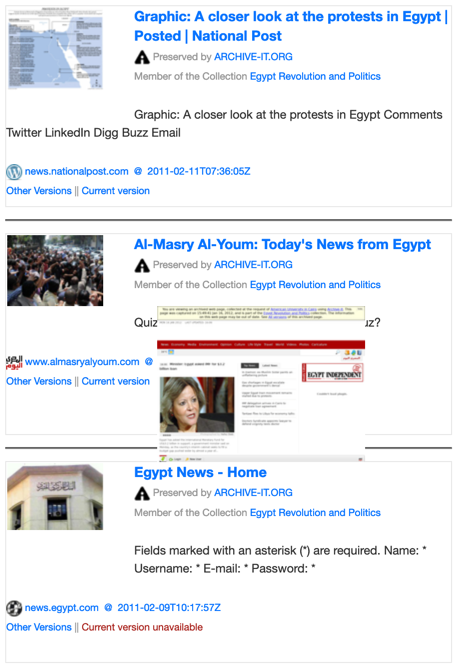}
\caption{A screenshot of part of a story constructed from a sample of $16$ mementos drawn from the $80,484$ mementos found in collection \emph{Egypt Revolution and Politics} (ID 2358), visualized using sc\textasciicircum t surrogates.}
\label{fig:example_story_socialcardwiththumbonhover}
\end{figure*}

\begin{table*}[t]
\caption{The collections and stories used in this study}
\begin{tabular}{r|l|l|r|l|r|r|r}

\textbf{Collection}  & \textbf{Collection} & \textbf{Collected} & \textbf{Diversity of} & \textbf{Collection} & \textbf{Collection Size} & \textbf{Story} & \textbf{\% of Story} \\
\textbf{ID} & \textbf{Name} & \textbf{By} & \textbf{Original Resource} & \textbf{Lifespan} & \textbf{(\# of Mementos} & \textbf{Size} & \textbf{with Good} \\
& & & \textbf{Domain Names} & & \textbf{from Seeds)} & & \textbf{Surrogates} \\
\hline
694           & April 16     & VT: Crisis, Tragedy, etc.    & 0.8391           & 48 weeks   & 374         & 17         &  88.24\%                                        \\
& Archive & & & & &   \\
\hline
1784          & Earthquake                        & IA Global Events             & 0.7656           & 9 weeks    & 1,080         & 27  & 85.19\%                                        \\
&  in Haiti  & & & & &   \\
\hline
2017          & Wikileaks 2010 & IA Global Events              & 0.575            & 3 years    & 3,333         & 24         & 70.83\%  \\
& Document Release  & & & & &  \\
& Collection & & & & &  \\
\hline
2358          & Egypt Revolution              & American University  & 0.2585           & 7 years   & 80,484          & 16         & 87.50\%    \\                                    
& and Politics & in Cairo  & & & & \\
\hline
2535          & Brazilian School                   & VT: Crisis, Tragedy, etc.     & 0.2604           & 5 days    & 1,540          & 26    & 73.08\%                                        \\
& Shooting & & & & &  \\
\hline
2823          & Russia Plane Crash    & VT: Crisis, Tragedy, etc.    & 0.7843           & 1 week     & 603         & 27 & 77.78\%                                        \\
& Sept 7,2011 & & & & &  \\
\hline
2950          & Occupy Movement                  & IA Global Events             & 0.5585           & 44 weeks    & 31,863        & 15 & 100.00\%                                       \\
& 2011/2012 & & & & &  \\
\hline
3649          & 2013 Boston                & IA Global Events             & 0.3766           & 1.9 years   & 2,421        & 27              & 96.30\%                                        \\
&  Marathon Bombing & & &  & & \\
\hline
3936          & United States          & IA Global Events             & 0.1177           & 4 years     & 24,583        & 16           & 93.75\%                                        \\
& Government & & & & & \\
& Shutdowns & & & & &  \\
\hline
4887          & Global Health Events            & NLM       & 0.2723           & 4 years    & 9,204         & 35 & 100.00\% \\
& web archive & & & & & \\
\end{tabular}

\label{tab:dataset}
\end{table*}

As a source of stories to display to the participants, we selected four stories from AlNoamany's 2016 dataset \cite{AlNoamany2017data}. Each story consists of ordered URI-Ms selected by a human curator to describe their collection. Details of the full dataset are shown in Table \ref{tab:dataset}. Some collections have mementos that are no longer available, possibly because they were removed by the curator. Some collections also have mementos that produce poor quality thumbnails. If a thumbnail failed to contain at least a heading describing some of the content within the memento, we considered it to be of poor quality. The last column in this table lists the percentage of the story that produced good quality surrogates.

We did not select collection \emph{Global Health Events web archive} (ID 4887) or \emph{United States Government Shutdowns} (ID 3936) because they have been repurposed to suit a larger topic and hence their 2016 stories no longer accurately reflect their content. Our four selections represent a variety of structural and semantic considerations. \emph{Occupy Movement 2011/2012} (ID 2950) was selected because it produces the best quality thumbnails. \emph{April 16 Archive} (ID 694) has the highest diversity of original resource domain names in its URIs \cite{Jones_Nelson_Weigle_2018_2}. \emph{Egypt Revolution and Politics}  (ID 2358) is a collection that is still currently being maintained and hence is the longest lived collection in the set. Collection \emph{Russia Plane Crash Sept 7,2011} (ID 2823) is about an event that is likely not familiar to American MT participants.

To compare against the as-is interface at Archive-It, we generated a facsimile of the Archive-It surrogates using Archive-It's stylesheets as well as metadata gathered using aiu \cite{jones2018aiu}. An example story using the Archive-It Facsimile surrogate is shown in Figure \ref{fig:example_story_archiveitlike}.

We employed MementoEmbed to generate a visualization of each story represented as thumbnails (Figure \ref{fig:example_story_thumbnails}) and again as social cards (Figure \ref{fig:example_story_socialcard}). From these, we developed three additional surrogate types combining social cards and thumbnails in order to see if a combination of the two produces better results. The surrogates for the story in Figure \ref{fig:example_story_socialcardthumbnail}, noted in this paper as sc+t, display the thumbnail to the right of the existing social card.  To produce the surrogates for the story in Figure \ref{fig:example_story_socialcardwiththumbasimg}, noted as sc/t, we replace the social card's striking image with the thumbnail. To conserve space and utilize interactivity, we use JavaScript in the surrogate shown in Figure \ref{fig:example_story_socialcardwiththumbonhover}, noted as sc\textasciicircum t, to allow the user to display the thumbnail if they hover their mouse over the striking image. All visualizations represented the same URI-Ms in the same order. 


The first page presented in the survey gave the participant instructions, indicating that they would view a story for 30 seconds and then be asked a question. We informed each participant that others were not necessarily all viewing the same visualization and that it might respond to mouse hovers, clicks, and other interactions. We did not provide any specific instruction on how to interact with the visualization beyond this. Once the participant had clicked through the instructions, the survey presented them the story. We recorded the initial timestamp of the story page load. The survey system used this timestamp to ensure that the participant was given 30 seconds to view the story. If the participant reloaded the page, the survey system recalculated the amount of time left, preventing them from gaining more time to view the story. We employed an HTML META tag to refresh the page after this time had expired. To address another avenue of potentially avoiding the 30 second timeout, we included JavaScript to prevent users from revisiting the story from their browser cache. Once the 30 seconds had expired, the participant was presented with a question.

The question consisted of checkboxes next to six new surrogates of the same type as the story that they just viewed. Participants were allowed as much time as possible to answer the question. We instructed the participant to select the two mementos that were drawn from the same collection that they had just viewed. We randomly generated the order of these surrogates, but we kept the same order for each collection. Our primary goal was to record how long the users took to answer each question, expecting them to find the two correct answers in all cases. In addition to instructing users to only select two responses, we also included JavaScript that prevented the user from selecting more or fewer than two. Our question follows Kittur's MT advice to use explicit, verifiable questions as part of the task \cite{kittur2008}. The simplicity of our question also avoids user fatigue \cite{kelly_methods_2007}. 

To produce the two correct answers, we randomly selected two URI-Ms from the same collection as the story shown to the participants. In choosing these URI-Ms, we discarded ones that used the same original resource domain as any memento in the story, avoiding issues where simple banners or logos might indicate that they are from the same collection. 

\begin{table*}[t]
\caption{Jaccard Distance of Named Entities between the different collections in the dataset. Blue indicates the collection that is most distant from the corresponding collection on the left. Green indicates the collection second most distant.}
\label{tab:jaccardne}

\begin{tabular}{ l ||  l | l | l | l | l | l | l | l | l | l}
 \textbf{Archive-It Collection} & \textbf{694} & \textbf{1784} & \textbf{2017} & \textbf{2358} & \textbf{2535} & \textbf{2823} & \textbf{2950} & \textbf{3649} & \textbf{3936} & \textbf{4887} \\
\hline \hline
694 -- April 16 Archive & 0.000 & 0.969 & 0.970 & \cellcolor{green!25} 0.981 & 0.961 & 0.968 & \cellcolor{blue!25}0.986 & 0.962 & 0.978 & 0.974 \\ \hline
1784 -- Earthquake in Haiti & 0.969 & 0.000 & 0.959 & 0.971 & 0.960 & \cellcolor{green!25} 0.975 & \cellcolor{blue!25}0.983 & 0.967 & 0.972 & 0.961 \\ \hline
2017 -- Wikileaks 2010 Document Release Collection & \cellcolor{green!25} 0.970 & 0.959 & 0.000 & 0.962 & 0.953 & \cellcolor{blue!25}0.977 & 0.965 & 0.959 & 0.956 & 0.966 \\ \hline
2358 -- Egypt Revolution and Politics & \cellcolor{green!25} 0.981 & 0.971 & 0.962 & 0.000 & 0.958 & \cellcolor{blue!25}0.985 & 0.965 & 0.971 & 0.955 & 0.970 \\ \hline
2535 -- Brazilian School Shooting & 0.961 & 0.960 & 0.953 & 0.958 & 0.000 & \cellcolor{blue!25}0.974 & \cellcolor{green!25} 0.967 & 0.955 & 0.952 & 0.961 \\ \hline
2823 -- Russia Plane Crash Sept 7,2011 & 0.968 & 0.975 & 0.977 & 0.985 & 0.974 & 0.000 & \cellcolor{blue!25}0.992 & 0.978 & \cellcolor{green!25} 0.987 & 0.977 \\ \hline
2950 -- Occupy Movement 2011/2012 & \cellcolor{green!25} 0.986 & 0.983 & 0.965 & 0.965 & 0.967 & \cellcolor{blue!25}0.992 & 0.000 & 0.974 & 0.942 & 0.981 \\ \hline
3649 -- 2013 Boston Marathon Bombing & 0.962 & 0.967 & 0.959 & 0.971 & 0.955 & \cellcolor{blue!25}0.978 & \cellcolor{green!25} 0.974 & 0.000 & 0.961 & 0.968 \\ \hline
3936 -- United States Government Shutdowns & \cellcolor{green!25} 0.978 & 0.972 & 0.956 & 0.955 & 0.952 & \cellcolor{blue!25}0.987 & 0.942 & 0.961 & 0.000 & 0.966 \\ \hline
4887 -- Global Health Events web archive & 0.974 & 0.961 & 0.966 & 0.970 & 0.961 & \cellcolor{green!25} 0.977 & \cellcolor{blue!25}0.981 & 0.968 & 0.966 & 0.000 \\ \hline
\end{tabular}

\end{table*}

\begin{figure*}[t]
	\centering

    \includegraphics[width=0.7\textwidth]{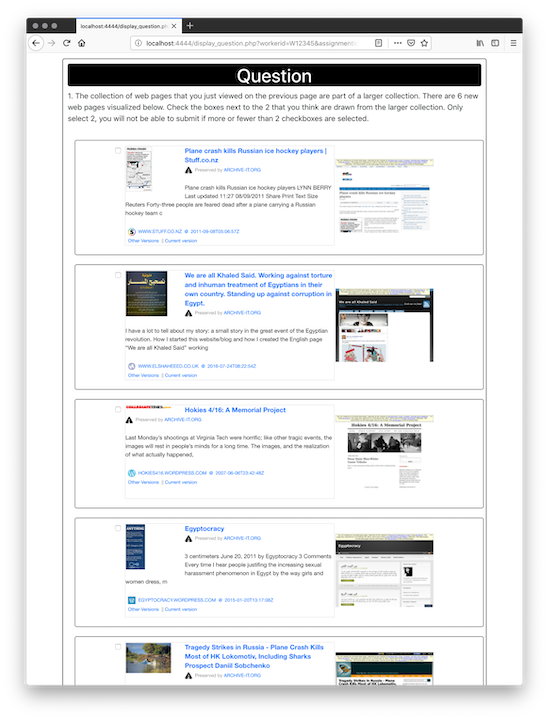}
		\caption{A screenshot of the question part of our survey for the sc+t surrogate for \emph{Egypt Revolution and Politics}.}
    \label{fig:question-screenshot}

\end{figure*}

To produce the four incorrect answers, we selected four other URI-Ms from semantically different collections. To determine which collections were semantically different from our story collection, we extracted entities from each collection in AlNoamany's dataset using Stanford NLP \cite{manning-EtAl:2014:P14-5}. We then computed the Jaccard distance between these entity sets and selected two collections with the greatest distance from our story collection. We randomly selected two URI-Ms from the most distant and second most distant collections. The distances between these collections are shown in Table \ref{tab:jaccardne}, where blue indicates the collections most distant from each other, and light green indicates second most distant. For the question for the \emph{Egypt} collection shown in Figure \ref{fig:question-screenshot}, the collection \emph{Russia Plane Crash Sept 7,2011} (ID 2823) has the greatest distance at 0.985. With a distance of 0.981, \emph{April 16 Archive} (ID 694) comes in second. Hence, two mementos are selected from each of these collections for the incorrect answers.

In all cases, we discarded URI-Ms that produced poor quality thumbnails to ensure that the quality of the memento did not affect the participant's choice. We also discarded URI-Ms that were off-topic, such as maintenance pages or 404 pages, as described in \cite{Jones_Nelson_Weigle_2018_1}. If a URI-M was discarded, we redrew to ensure that there were two selections from the collection they had just viewed, two selections from the most semantically distant collection, and two selections from the second most distant collection. We then randomly sorted the six URI-Ms and generated the surrogates. MementoEmbed cards contain the name of the collection from which they were selected.  To avoid giving an unfair advantage to social cards, the collection name was removed from the social cards used in the question. Appendix B contains screenshots of the questions shown to study participants.

In addition to gathering answers from the participant, we also collected information about their behavior. Our survey system recorded a timestamp for the load of the question page. It then recorded the timestamp for the load of the completion code. The time the participant took to answer the question is the difference between these two timestamps. We employed JavaScript to record all link clicks and hovers over images and links. This provides us with several data points with respect to those surrogates: the correctness of their answers, the time the user took to answer the question, and how they interacted with the story. We ran Student's t-test between all pairs of surrogates for completion times and the number of correct answers.

\begin{table*}[t]
\begin{tabular}{l||l|l|l|l|l||l|l|l|l|l}
& \multicolumn{5}{c||}{\textbf{Mean}} & \multicolumn{5}{c}{\textbf{Median}} \\
\textbf{Surrogate Type} & \textbf{694} & \textbf{2358} & \textbf{2823} & \textbf{2950} & \textbf{Overall} & \textbf{694} & \textbf{2358} & \textbf{2823} & \textbf{2950} & \textbf{Overall} \\
& (VATech) & (Egypt) & (Russia) & (Occupy) & & (VATech) & (Egypt) & (Russia) & (Occupy) & \\
\hline
Archive-It Facsimile & 100.27 & 412.93 & 36.77 & 48.14 & 149.53 & 42.07 & 34.71 & 25.64 & 32.20 & 33.46 \\
\hline
Browser Thumbnails & 68.66 & 272.54 & 59.52 & 44.15 & 111.22 & 53.76 & 103.11 & 36.67 & 43.42 & 53.30 \\
\hline
Social Cards & 52.91 & 56.11 & 34.10 & 41.37 & 46.12 & 40.28 & 37.68 & 32.87 & 44.24 & 35.89 \\
\hline
sc+t & 62.19 & 106.80 & 26.42 & 56.03 & 62.86 & 43.90 & 52.18 & 28.82 & 39.04 & 40.07 \\
\hline
sc/t & 48.01 & 130.89 & 47.81 & 42.69 & 67.35 & 38.34 & 32.93 & 38.87 & 38.41 & 38.38 \\
\hline
sc\textasciicircum t & 51.67 & 55.50 & 27.35 & 116.29 & 62.70 & 53.14 & 46.35 & 21.43 & 36.63 & 38.07 \\
\end{tabular}
\caption{Mean and median completion times for each surrogate type per collection and overall}
\label{tab:completion-times}
\end{table*}

\begin{table*}[t]
\begin{tabular}{l||l|l|l|l|l||l|l|l|l|l}
& \multicolumn{5}{c||}{\textbf{Mean}} & \multicolumn{5}{c}{\textbf{Median}} \\
\textbf{Surrogate Type} & \textbf{694} & \textbf{2358} & \textbf{2823} & \textbf{2950} & \textbf{Overall} & \textbf{694} & \textbf{2358} & \textbf{2823} & \textbf{2950} & \textbf{Overall} \\
& (VATech) & (Egypt) & (Russia) & (Occupy) & & (VATech) & (Egypt) & (Russia) & (Occupy) & \\
\hline
Archive-It Facsimile & 1.6 & 1.0 & 1.6 & 1.0 & 1.30 & 2.0 & 1.0 & 2.0 & 1.0 & 1.5 \\
\hline
Browser Thumbnails & 1.2 & 1.4 & 1.6 & 1.6 & 1.45 & 1.0 & 2.0 & 2.0 & 2.0 & 2.0 \\
\hline
Social Cards & 1.6 & 1.8 & 1.6 & 2.0 & 1.75 & 2.0 & 2.0 & 2.0 & 2.0 & 2.0 \\
\hline
sc+t & 1.2 & 1.8 & 1.8 & 2.0 & 1.70 & 1.0 & 2.0 & 2.0 & 2.0 & 2.0 \\
\hline
sc/t & 1.4 & 1.8 & 1.8 & 1.2 & 1.55 & 2.0 & 2.0 & 2.0 & 2.0 & 2.0 \\
\hline
sc\textasciicircum t & 2.0 & 1.2 & 2.0 & 1.6 & 1.70 & 2.0 & 2.0 & 2.0 & 2.0 & 2.0 \\
\end{tabular}
\caption{Mean and median completion correct answers for each surrogate type per collection and overall}
\label{tab:correct-answers}
\end{table*}

\subsection{Results}

Table \ref{tab:completion-times} displays the mean and median question completion times for each surrogate. At 149.53 seconds, the Archive-It Facsimile surrogates have the highest mean time for answering the question. Browser thumbnails come in second highest at 111.22 seconds. Social cards have the lowest overall mean at 46.12 seconds. The sc+t and sc\textasciicircum t have means slightly greater than 62 seconds. The sc/t surrogate comes in slightly higher at 62.86 seconds. We executed the Student's t-test on the times for all pairs of surrogates. No values are statistically significant at $p < 0.05$. Social cards compared to browser thumbnails produces the lowest $p$-value at $p=0.190$. The next lowest $p$-value is $p=0.202$ for social cards compared to the Archive-It Facsimile. In spite of the mean values, these p-values indicate that our results provide weak evidence that the Archive-It Facsimile or thumbnails take the most amount of time to evaluate or that social cards take less time. The medians demonstrate that some outliers are skewing these means. The Archive-It Facsimile has the lowest median at 33.46 seconds, followed by social cards at 35.89 seconds. The median completion time for browser thumbnails is highest at 53.30 seconds. The combinations of social card and thumbnail all have medians between 38 and 40 seconds. Thus, even though the browser thumbnails still have the highest median, the p-values still demonstrate that we have not established that thumbnails take longer to process.

Table \ref{tab:correct-answers} displays the mean and median number of correct answers for each surrogate. With only 2 correct answers out of 6, the distribution of potential values is small. Social cards score highest with a mean correct answer score of 1.75, followed by a tie between sc+t and sc\textasciicircum t at 1.70. The Archive-It Facsimile mean is the lowest at 1.30. The medians are 2.0 for all but the Archive-It Facsimile at 1.5. The Archive-It Facsimile paired with the social card comes closest to statistical significance at $p < 0.05$ with $p=0.0569$. The next lowest $p$-values are for Archive-It vs. sc+t at $p=0.0770$ and Archive-It vs. sc\textasciicircum t at $p=0.108$. Within collection 2358, social cards, sc+t, and sc/t all fare better than the Archive-It Facsimile at $p=0.0650$ in all cases. Within collection 2950, social cards and sc+t all fare better than the Archive-It Facsimile, both at $p=0.0560$. Familiarity with the topics of some collections may have influenced the results and this is why we had selected different collections for this evaluation. The close p-values indicate that our general results of social cards compared to the Archive-It surrogate are similar to those of Capra et al. \cite{capra2013}, even though Capra focuses on information retrieval and not summarization.

The variation in the quality Archive-It Facsimile surrogates may also have shaped the results. Some of the Archive-It surrogates in the story for the \emph{Egypt} collection contained as many as 12 additional metadata fields while others from the same collection were minimal Archive-It surrogates. Almost all of the surrogates in the story for the \emph{Occupy} collection contained only the additional metadata field \texttt{Group}. In those cases \texttt{Group} contained values like \texttt{Social Media} and \texttt{News Sites and Articles}, text that provides little information specific to the collection. In contrast, almost all Archive-It surrogates for stories from the \emph{Russia} and \emph{VATech} collections contained the additional title metadata field. For a story consisting of mostly minimal Archive-It surrogates, it is possible that a small number of metadata-rich surrogates provided enough information for the user to effectively answer the question. 

\begin{figure}
\includegraphics[width=0.5\textwidth]{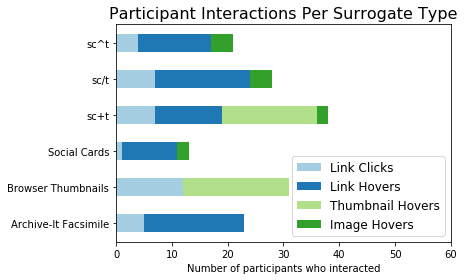}
\caption{The number of user interactions per surrogate type, broken down by image hovers, link hovers, and link clicks.}
\label{fig:user_interaction}
\end{figure}

Because each story has a different size, it is difficult to normalize the recorded user interactions across all stories. We chose to tally the number of users who hovered over images, hovered over links, and clicked links. The results are shown in Figure \ref{fig:user_interaction}. This engagement gives some insight into the amount of work each participant put into interacting with the story that they viewed. Recall that there are 20 total participants for each surrogate type.

Social cards produced interactions from the least number of participants. The most engagement with social cards involved link hovering.  We recorded 10 participants that hovered over links in social cards but only one participant clicked on links. Only two participants hovered over images. As noted above, they also spent the least amount of time with social cards when answering the question, and typically answered the questions more accurately. It is possible that they felt that social cards provided sufficient information for them to answer the question quickly and correctly.

The Archive-It surrogate has no images, and hence no image hovers. We recorded 18 participants hovering over the links, but only five participants clicked on these links. In spite of no images being present, they hovered and clicked more links than with social cards. Considering their performance in both time and correct answers, it is possible that they felt that more interaction was necessary to understand the story.

With browser thumbnails the image is the anchor of the link, hence every hover over an image is also a hover over a link. To account for this, we created a separate category named ``thumbnail hovers'' combining link and image hovers for thumbnails. Browser thumbnails experienced the most link clicks with 18 participants choosing to open the page behind the surrogate rather than just relying upon the surrogate alone. We recorded 19 participants hovering over thumbnails for this surrogate type. We did not measure if the user magnified each thumbnail or how long they viewed the pages that they opened. 

The sc+t consists of a social card with a thumbnail beside it, also of 201 pixels wide, meaning that the sc+t surrogate also supports thumbnail hovers. This surrogate type produced interactions from the most users. This level of engagement is surprising considering that the mean completion time for sc+t is shorter than that of browser thumbnails. We recorded 17 participants hovering over the thumbnail portion of the sc+t surrogate. This is two fewer participants than for the browser thumbnail surrogate, but still indicates a lot of mouse movement around thumbnails. Only two participants chose to hover over the non-thumbnail images on the social cards. This surrogate type did not inspire as much link clicking as browser thumbnails, with only seven participants clicking links. This is still higher than social cards alone, where only one participant clicked links.

The sc/t surrogate contains a thumbnail instead of the striking image normally found in social cards, so the image hovers are actually over thumbnails. We do not count them as thumbnail hovers because these images are not also anchors for links. For sc/t, four participants hovered over images, 17 participants hovered over links, and seven participants clicked links on these surrogates. This difference in behavior, coupled with the different response times and accuracy for sc/t compared to social cards suggests that including the thumbnail rather than a striking image drawn from the page may inspire more activity on the part of the user.

The sc\textasciicircum t surrogate provides a thumbnail if the user hovers over the striking image. Only four participants actually discovered this capability. In addition, 13 participants hovered over links, and seven participants clicked links.


Social cards inspired the least user interactions and the least link clicks. Perhaps the social card inspired more confidence and fewer participants needed to view the pages behind them. In contrast, the most users clicked on thumbnails to open links. Perhaps they found the thumbnails harder to read and felt less confident about their content. The most participants interacted with the sc+t surrogate in some way. More link clicks occurred in all cases where thumbnails were present. This difference in behavior, coupled with the different response times and accuracy for sc/t compared to social cards suggests that including the thumbnail rather than a striking image drawn from the page may inspire more activity on the part of the user. It is possible that our survey measured users zooming in on thumbnails to see them better. Link hovers have a strong correlation with completion time at Pearson's $r = 0.562$, but other interactions, including link clicks, had much weaker correlations to completion time at $|r| > 0.20$. Link hovers have a weak negative correlation with answer accuracy at $r = -0.298$, but other interactions had much weaker correlations to accuracy at $|r| > 0.20$. It is possible that participants hovered over links to read the URLs in their browser status bar before making their choice.

\section{Future Work}

In our previous work \cite{Jones_Nelson_Weigle_2018_2}, we organized Archive-It collections into four categories. The collections in this study fit into the category of type Time Bounded - Spontaneous. AlNoamany et al. \cite{alnoamany2017} discuss different types of stories that can be derived from web archive collections. All of the stories used in this study are of the type sliding page, sliding time. A study examining if some surrogates perform better for other types of collections and other types of stories may be beneficial.

The type of question asked of the participant may also allow us to determine which aspects of these surrogates work best for different purposes. For example, if we present the participant with a series of images drawn from various collections, it may indicate how well images function for understanding.

How well do the contents of the surrogates compare to the underlying documents they visualize? Computing the overlap between the text present in the surrogate and the information of the documents they visualize may provide a measure of how well a surrogate is expected to perform. These results can be contrasted with how well users actually perform.

Our results are similar to those observed by Capra et al. \cite{capra2013}. What other visualization elements from search engine result pages may be useful to our summarization efforts? Perhaps we should next explore concepts like entity cards \cite{Bota:2016:PYC:2854946.2854967} which summarize multiple resources from a collection that center on a specific entity.

Another area of interest to explore may be the sources of content. If users can identify sources via domain name on the social cards, full URI in the Archive-It surrogate, or recognizing layouts and logos in the browser thumbnail, then it may affect how they view the content of the story and hence the underlying collection.

Do users visually scan differently for thumbnails vs. social cards or the Archive-It like interface? Perhaps techniques like eye tracking can be introduced to evaluate their behavior to ensure that information is presented in a location optimized for their behavior.

Further measuring different interactions with other parts of the surrogate may offer additional insight. We assume that users are zooming in to better view thumbnails, but we have no way of measuring that at this time.

Determining the most effective visualization is only one important part of our work. The stories in this study were generated by human curators. We are also building on the work of AlNoamany et al. \cite{alnoamany2017} by creating new algorithms to automatically select mementos that best represent the collection.

\section{Conclusions}

Surrogates have been used in the past to answer the question of ``should I click on this?'' In this work, we instead consider the use of surrogates in a group to answer the question ``What does the underlying collection contain?'' We examined the variation in metadata present in Archive-It surrogates and found that, in spite of more than half of Archive-It surrogates missing data, information could still potentially be gleaned from the URL present in a minimal surrogate. We asked participants from MT to view a story visualized using a given surrogate. We then gave them a question with six mementos visualized using the same surrogate and asked them to choose the two from the six that they believed belonged to the same collection as the story that they just viewed. The type of surrogate does not influence the time to complete the task, but social cards and social cards side-by-side with thumbnails \textit{probably} provide better collection understanding than the existing Archive-It interface at $p=0.0569$, and $p=0.0770$, respectively. This is consistent with results from a study by Capra et al. \cite{capra2013} comparing the performance of social cards to text snippets in search results.

We also found that user interactions differ between surrogate types, with social cards having the fewest participants interact and a combination of social card side-by-side with thumbnail encouraging the most participants to interact. Because participants also appear to hover and click more when thumbnails are present, we postulate that users engage more with browser thumbnails than other surrogate elements, possibly to zoom in and see details. 

For collection summarization, the overall goal of surrogates is to convey aboutness without requiring the user to click on the underlying link. In this case, social cards appear to require less interaction, provide higher accuracy, and allow the users to answer our question in less time. These results are encouraging for users of social cards. Social cards require fewer resources to generate and store than thumbnails. Archive-It surrogates require humans to construct metadata, but social cards can be generated dynamically from existing web page content. Users also appear to interact with social cards less, possibly indicating that they find them easier to use. These features indicate that social cards may be the best surrogate for use in summarizing web archive collections, displaying stories on live web curation platforms, viewing saved items in bookmarking applications, sharing on social media, and beyond.

\begin{acks}
This work has been supported in part by the Institute of Museum and Library Services (LG-71-15-0077-15).
\end{acks}

\bibliographystyle{ACM-Reference-Format}
\bibliography{references}

\clearpage

\appendix

\section{The Current Collection Understanding Process for Archive-It Collections}

\begin{figure*}

		\includegraphics[width=\textwidth]{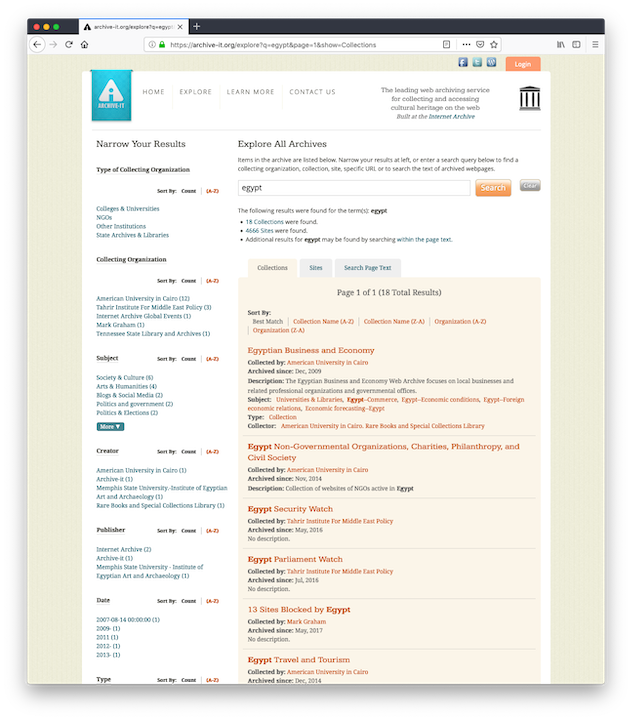}
		\caption{Step 1: Form a query to find the collection desired.}
		\label{fig:manual_collection_review_step1}
\end{figure*}

\begin{figure*}
		\includegraphics[width=\textwidth]{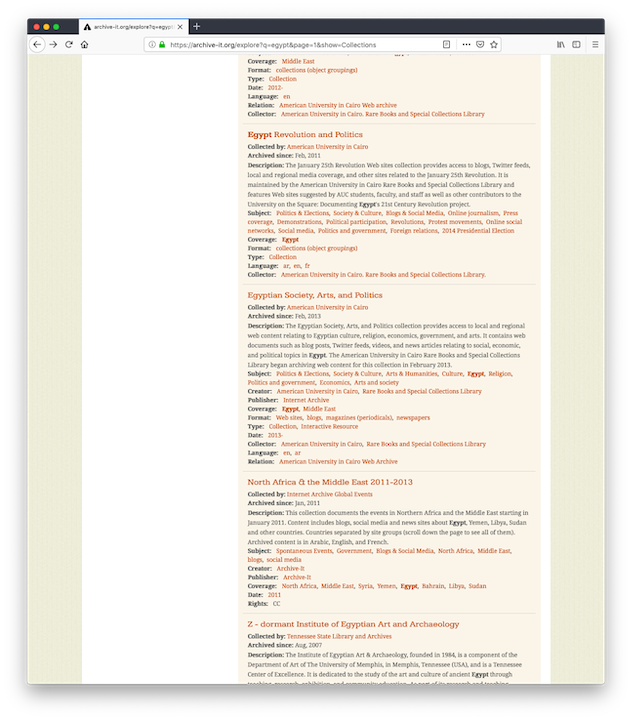}
		\caption{Step 2: Choose a collection from the list that may meet the information need.}
		\label{fig:manual_collection_review_step2}
\end{figure*}

\begin{figure*}

		\includegraphics[width=\textwidth]{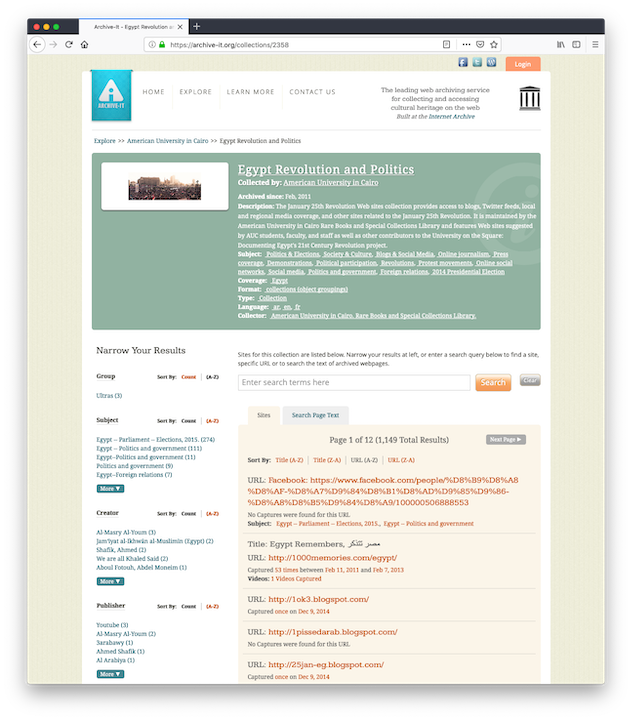}
		\caption{Step 3: View the collection page for the collection.}
		\label{fig:manual_collection_review_step3}
\end{figure*}

\begin{figure*}

		\includegraphics[width=\textwidth]{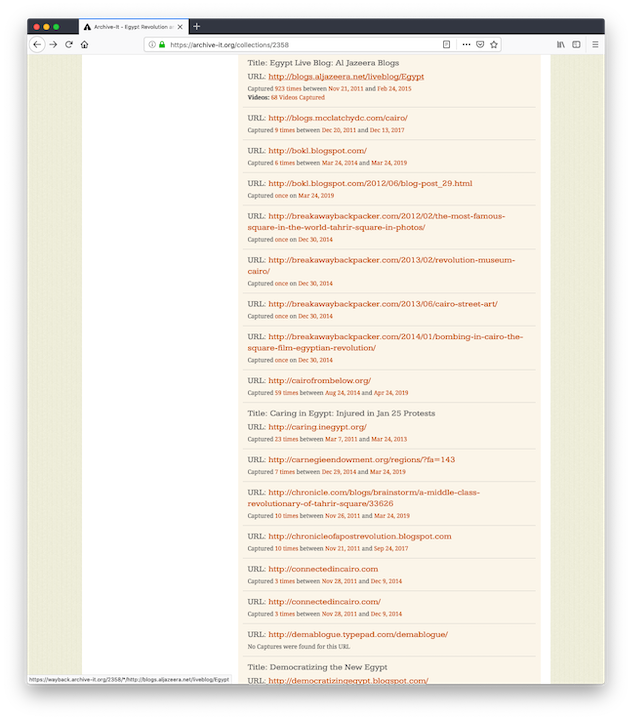}
		\caption{Step 4: Choose a seed from the list that may meet the information need. Collection \emph{Egypt Revolution and Politics} contains 1,149 seeds.}
		\label{fig:manual_collection_review_step4}
%
	
%
\end{figure*}

\begin{figure*}

		\includegraphics[width=\textwidth]{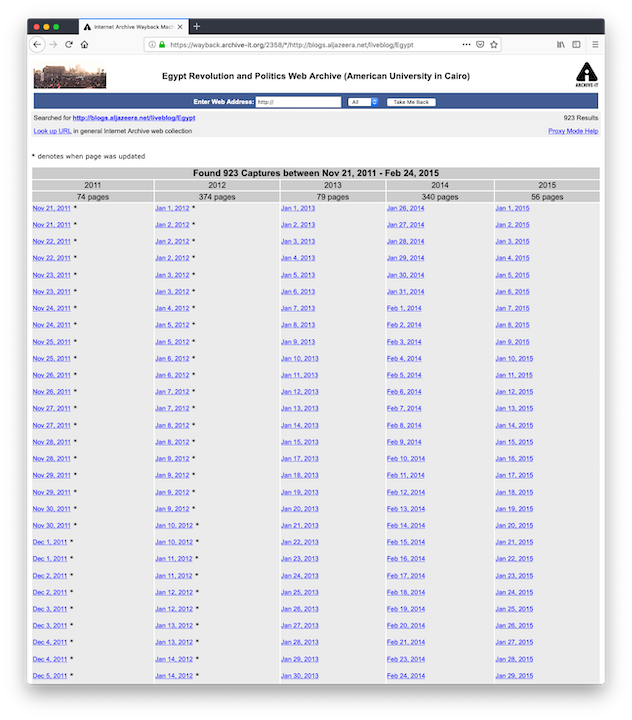}
		\caption{Step 5: Choose a memento of that seed to examine. This seed has was used in 923 crawls, producing 923 mementos. Recall that these mementos are just the start of the crawl and that there may be more linked from them.}
		\label{fig:manual_collection_review_step5}
\end{figure*}

\begin{figure*}

		\includegraphics[width=\textwidth]{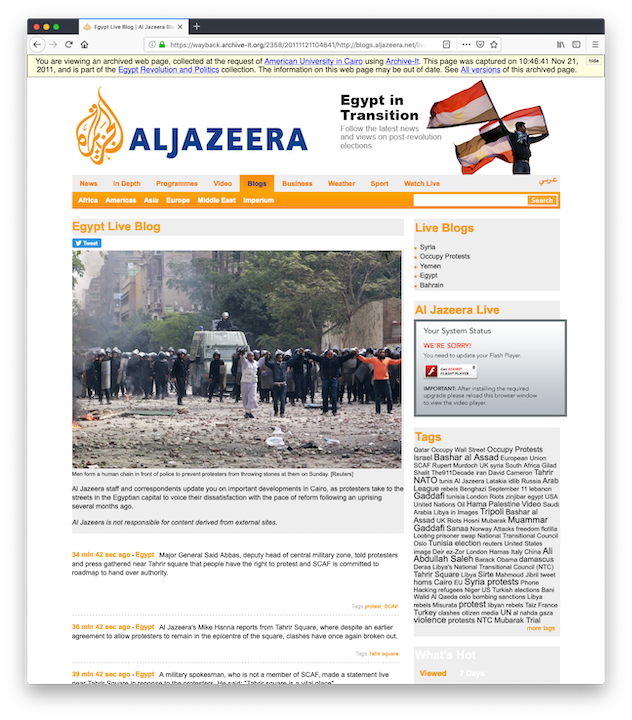}
		\caption{Step 6: Read the text of the memento to learn about its contents.}
		\label{fig:manual_collection_review_step6}
%
\end{figure*}

\begin{figure*}

		\includegraphics[width=\textwidth]{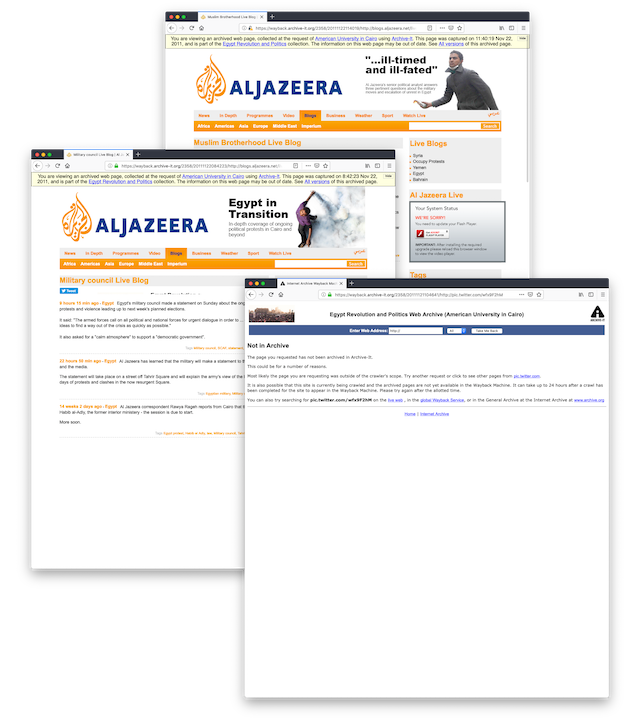}
		\caption{Step 7: View mementos linked from that memento, and mementos linked from those mementos until we reach links to pages that the crawler did not preserve.}
		\label{fig:manual_collection_review_step7}
\end{figure*}

\begin{figure*}

		\includegraphics[width=\textwidth]{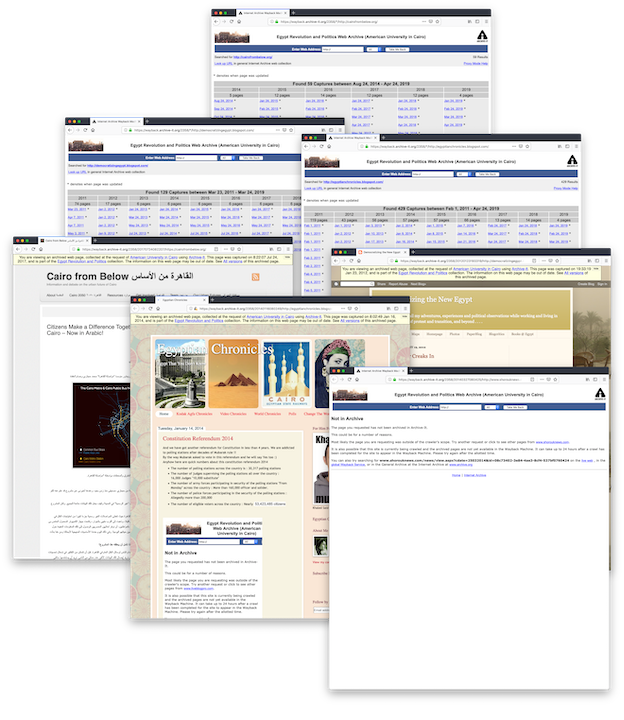}
		\caption{Step 8: Repeat steps 4-7 until enough information about the collection has been amassed to determine that it meets the information need. We will iterate back through seeds, mementos, and broken links, reviewing many documents in the process. If the collection is not meeting the information need, we may need to restart at step 1.}
		\label{fig:manual_collection_review_step8}
%

\end{figure*}

Understanding an Archive-It collection is an iterative, tedious process. Figures \ref{fig:manual_collection_review_step1} through \ref{fig:manual_collection_review_step8} provide the steps necessary to manually achieve  collection understanding for an Archive-It collection. To begin at step 1 (Figure \ref{fig:manual_collection_review_step1}), a user must first have a query. As seen in the search results from the screenshot in Figure \ref{fig:manual_collection_review_step1}, not all collections contain metadata, thus we are often left with their collection title to make a decision.  In step 2 (Figure \ref{fig:manual_collection_review_step2}), we choose a collection from the list that we think will meet our information need. In step 3 (Figure \ref{fig:manual_collection_review_step3}), we view the collection, navigating through its seeds and choosing one in step 4 (Figure \ref{fig:manual_collection_review_step4}). Note how not all seeds have metadata. Once we have chosen a seed, we view its mementos in step 5 (Figure \ref{fig:manual_collection_review_step5}) and choose one. Note how this interface provides the dates for each memento, but no other information. In this example there are 923 mementos for this seed and this seed was one of 1,149 seeds in this collection. This first memento, however, is just the start of this crawl and other mementos were captured that were linked from that page, hence in Step 7 (Figure \ref{fig:manual_collection_review_step7}), we review the linked pages until we reach an Archive-It error page indicating that the linked page that was not crawled. At this point, we understand the contents of a single crawl of a single seed of a single collection. To understand the rest of the collection, we must review other seeds and their mementos, and their linked mementos (Figure \ref{fig:manual_collection_review_step8}). 

If this collection meets our information needs then we just need to iterate from the seed level. If this collection is not meeting our information need, then we have two options. We can restart at step 2 by choosing one of the other 17 collections that matched our search term. The first two collections in the list have 95 and 331 seeds to review, respectively. Alternatively, if we believe that our search terms are not successful, then we must restart at step 1 by reformulating our query.

\clearpage
\section{Screenshots Of Pages Shown to Study Participants}

The following sections display the task instructions, questions, and an example completion code page shown to the study participants. Due to space limitations, we were unable to include screenshots of the stories themselves.

\subsection{Task Instructions}

\begin{figure*}[htbp!]
\centering
\includegraphics[width=\textwidth]{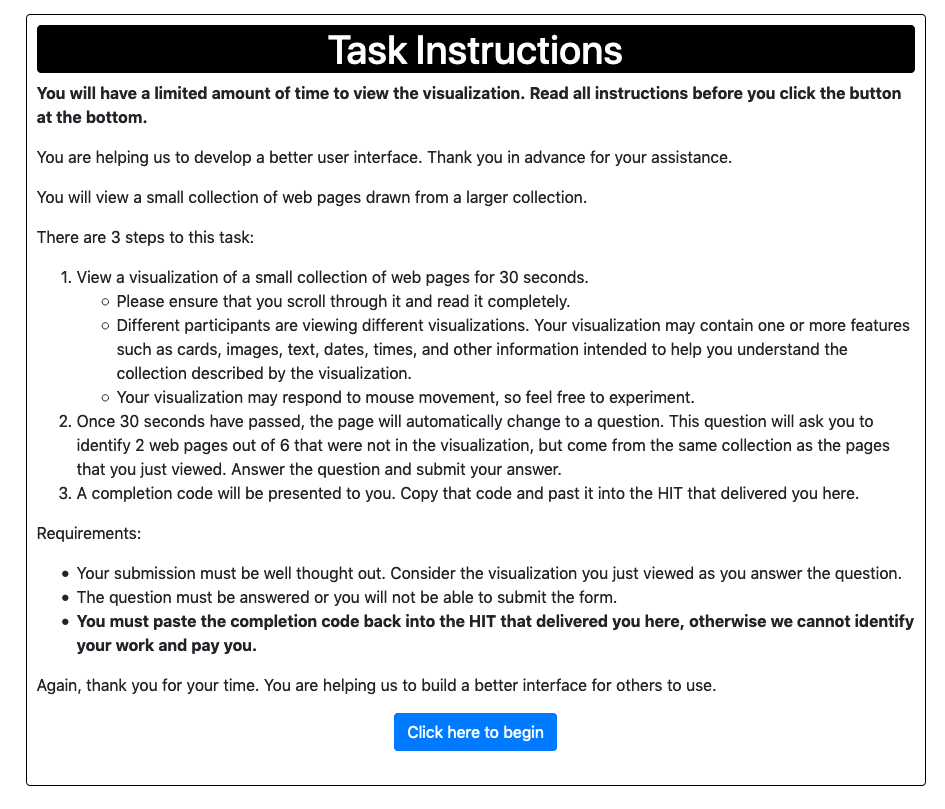}
\caption{Screenshot of the task instructions given to study participants}
\end{figure*}

\clearpage
\subsection{Questions}

\begin{figure*}[htbp]
\includegraphics[height=8in]{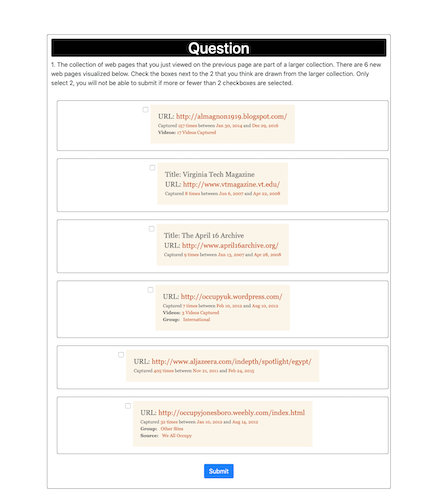}
\caption{Screenshot of question using Archive-It facsimile surrogates for \emph{April 16 Archive}}
\end{figure*}

\begin{figure*}[htbp]
\includegraphics[height=8in]{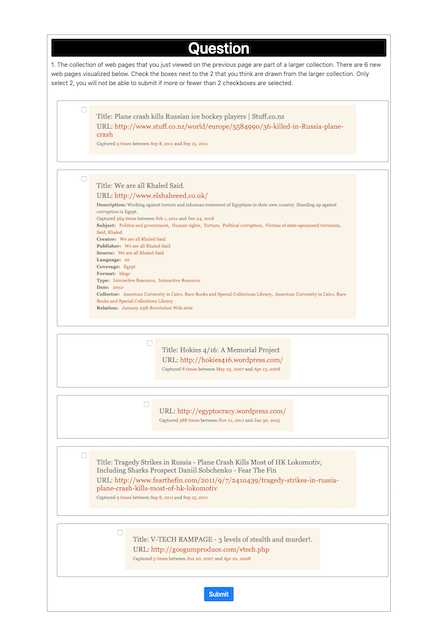}
\caption{Screenshot of question using Archive-It facsimile surrogates for \emph{Egypt Revolution and Politics}}
\end{figure*}

\begin{figure*}[htbp]
\includegraphics[height=8in]{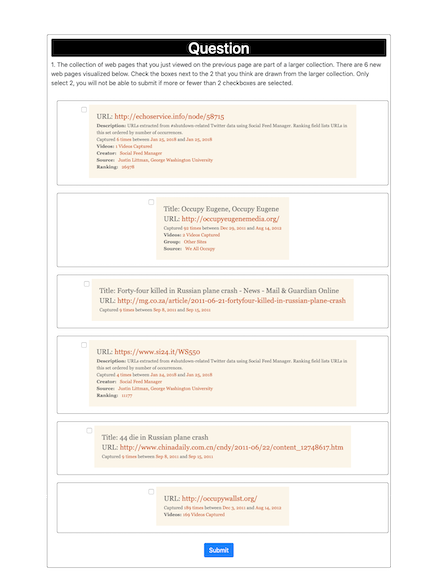}
\caption{Screenshot of question using Archive-It facsimile surrogates for \emph{Russia Plane Crash Sept 7,2011}}
\end{figure*}

\begin{figure*}[htbp]
\includegraphics[height=8in]{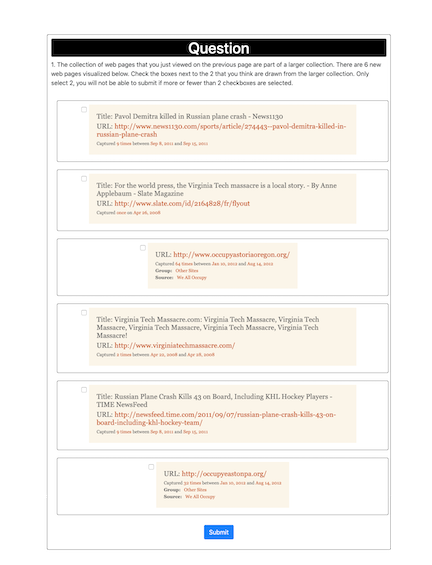}
\caption{Screenshot of question using Archive-It facsimile surrogates for \emph{Occupy Movement 2011/2012}}
\end{figure*}

\begin{figure*}[htbp]
\includegraphics[height=8in]{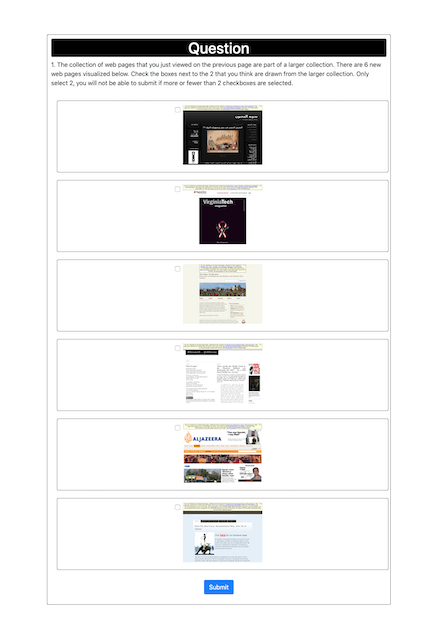}
\caption{Screenshot of question using browser thumbnail surrogates for \emph{April 16 Archive}}
\end{figure*}

\begin{figure*}[htbp]
\includegraphics[height=8in]{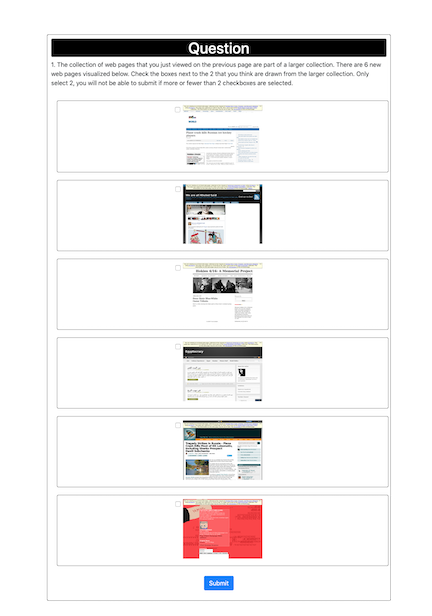}
\caption{Screenshot of question using browser thumbnail surrogates for \emph{Egypt Revolution and Politics}}
\end{figure*}

\begin{figure*}[htbp]
\includegraphics[height=8in]{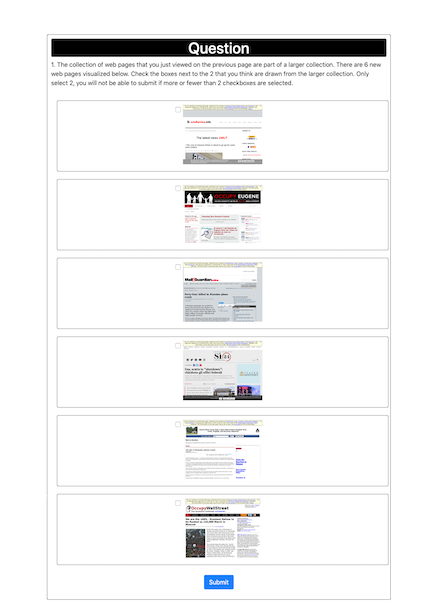}
\caption{Screenshot of question using browser thumbnail surrogates for \emph{Russia Plane Crash Sept 7,2011}}
\end{figure*}

\begin{figure*}[htbp]
\includegraphics[height=8in]{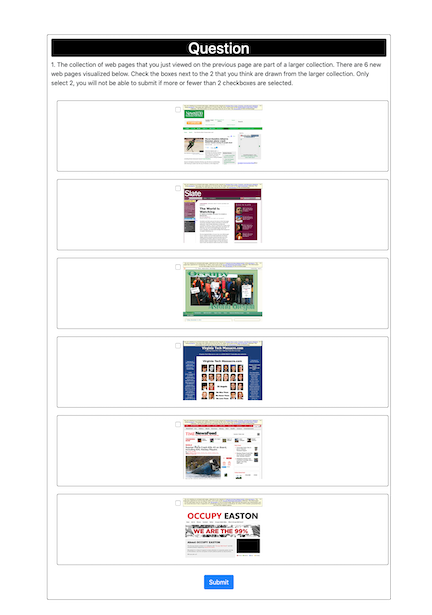}
\caption{Screenshot of question using browser thumbnail surrogates for \emph{Occupy Movement 2011/2012}}
\end{figure*}

\begin{figure*}[htbp]
\includegraphics[height=8in]{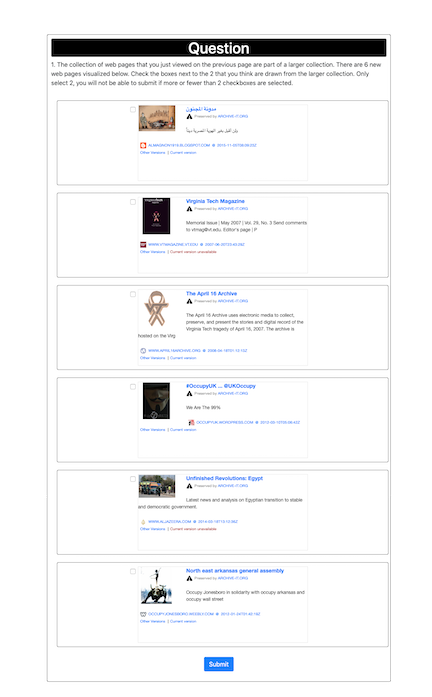}
\caption{Screenshot of question using social card surrogates for \emph{April 16 Archive}}
\end{figure*}

\begin{figure*}[htbp]
\includegraphics[height=8in]{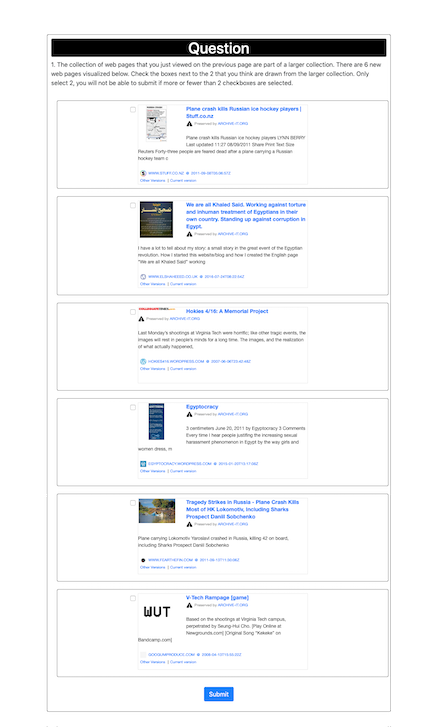}
\caption{Screenshot of question using social card surrogates for \emph{Egypt Revolution and Politics}}
\end{figure*}

\begin{figure*}[htbp]
\includegraphics[height=8in]{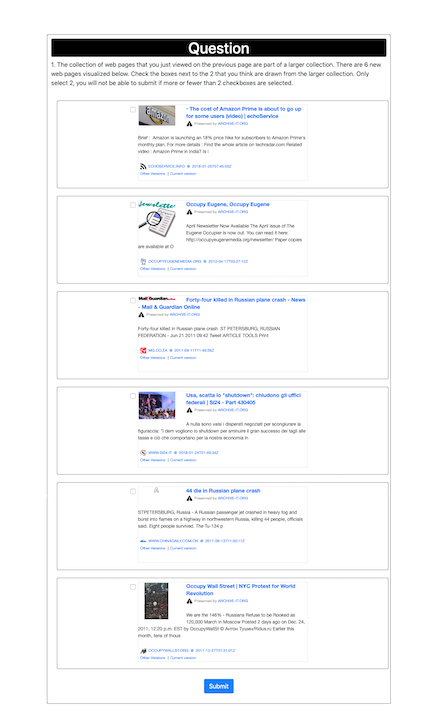}
\caption{Screenshot of question using social card surrogates for \emph{Russia Plane Crash Sept 7,2011}}
\end{figure*}

\begin{figure*}[htbp]
\includegraphics[height=8in]{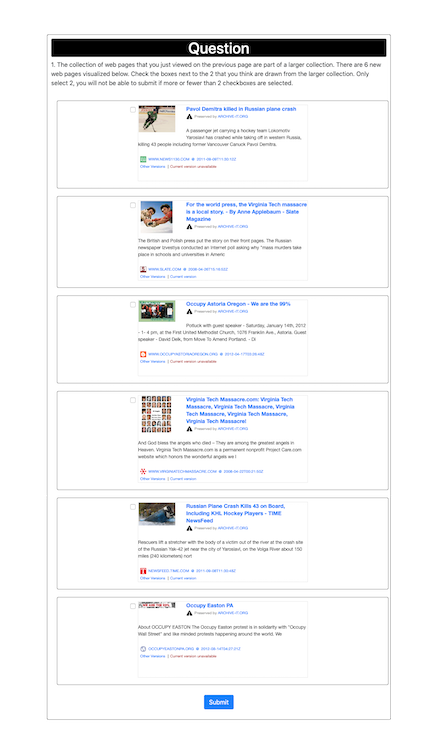}
\caption{Screenshot of question using social card surrogates for \emph{Occupy Movement 2011/2012}}
\end{figure*}

\begin{figure*}[htbp]
\includegraphics[height=8in]{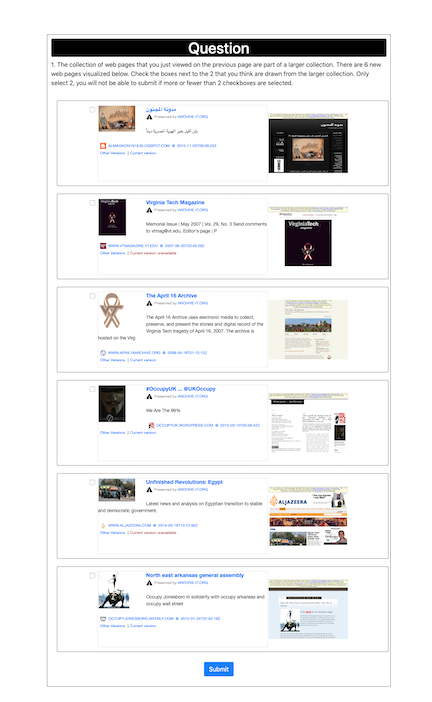}
\caption{Screenshot of question using sc+t surrogates for \emph{April 16 Archive}}
\end{figure*}

\begin{figure*}[htbp]
\includegraphics[height=8in]{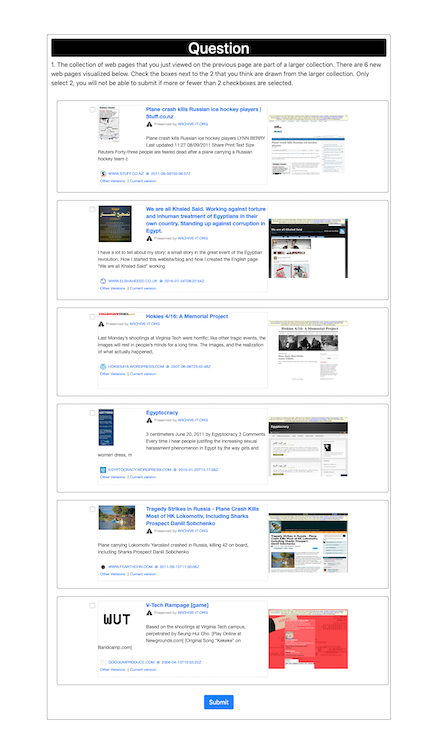}
\caption{Screenshot of question using sc+t surrogates for \emph{Egypt Revolution and Politics}}
\end{figure*}

\begin{figure*}[htbp]
\includegraphics[height=8in]{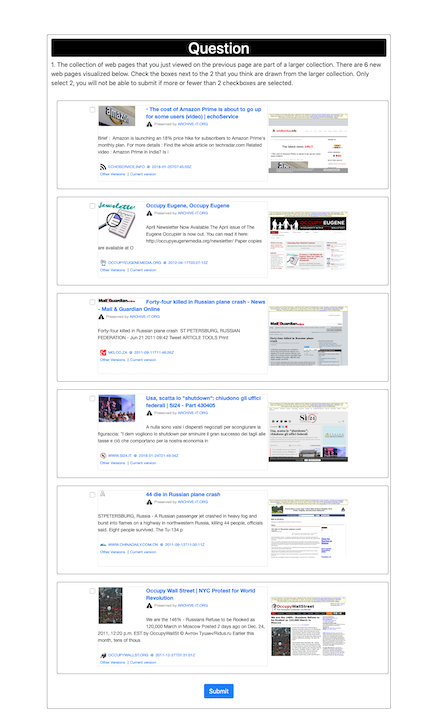}
\caption{Screenshot of question using sc+t surrogates for \emph{Russia Plane Crash Sept 7,2011}}
\end{figure*}

\begin{figure*}[htbp]
\includegraphics[height=8in]{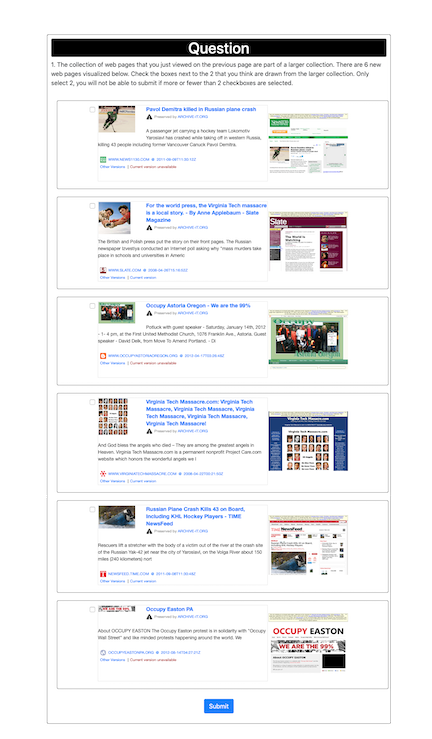}
\caption{Screenshot of question using sc+t surrogates for \emph{Occupy Movement 2011/2012}}
\end{figure*}

\begin{figure*}[htbp]
\includegraphics[height=8in]{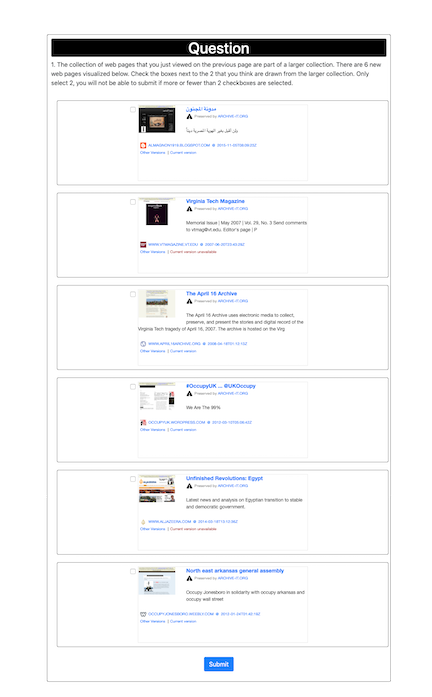}
\caption{Screenshot of question using sc/t surrogates for \emph{April 16 Archive}}
\end{figure*}

\begin{figure*}[htbp]
\includegraphics[height=8in]{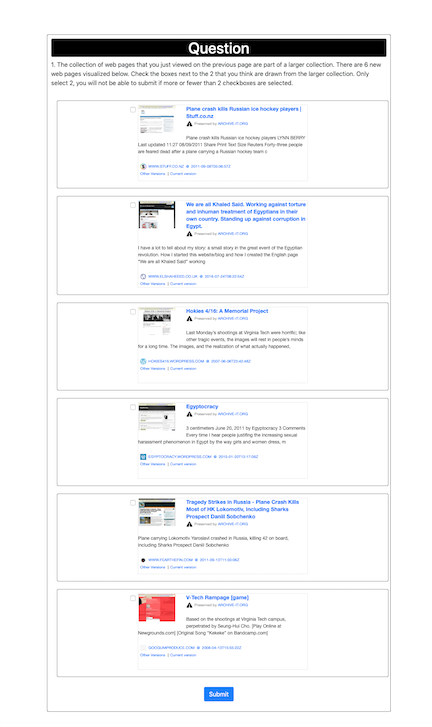}
\caption{Screenshot of question using sc/t surrogates for \emph{Egypt Revolution and Politics}}
\end{figure*}

\begin{figure*}[htbp]
\includegraphics[height=8in]{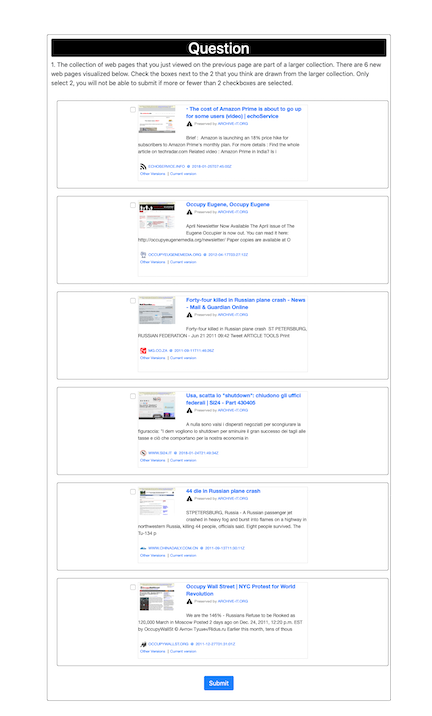}
\caption{Screenshot of question using sc/t surrogates for \emph{Russia Plane Crash Sept 7,2011}}
\end{figure*}

\begin{figure*}[htbp]
\includegraphics[height=8in]{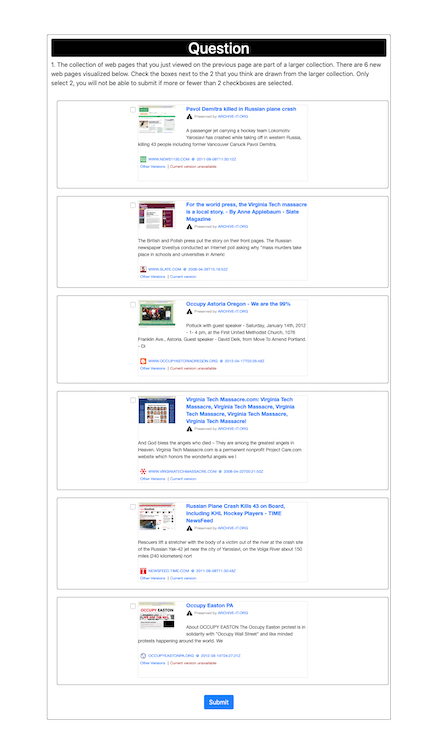}
\caption{Screenshot of question using sc/t surrogates for \emph{Occupy Movement 2011/2012}}
\end{figure*}

\begin{figure*}[htbp]
\includegraphics[height=8in]{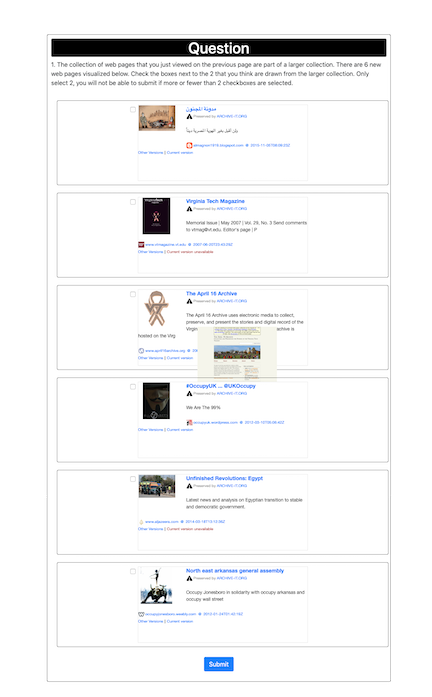}
\caption{Screenshot of question using sc\textasciicircum t surrogates for \emph{April 16 Archive}}
\end{figure*}

\begin{figure*}[htbp]
\includegraphics[height=8in]{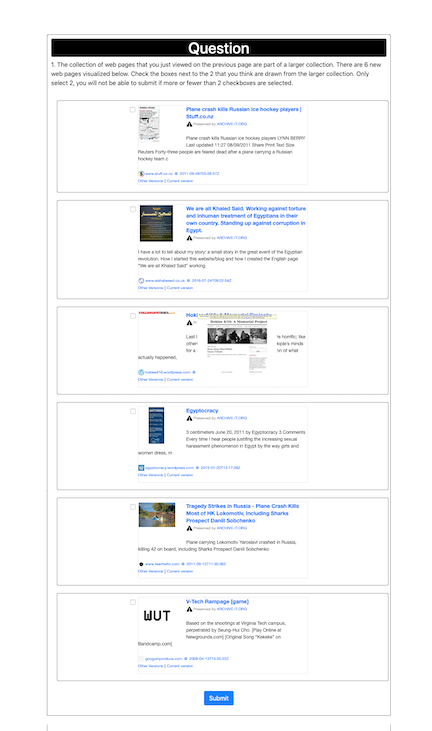}
\caption{Screenshot of question using sc\textasciicircum t surrogates for \emph{Egypt Revolution and Politics}}
\end{figure*}

\begin{figure*}[htbp]
\includegraphics[height=8in]{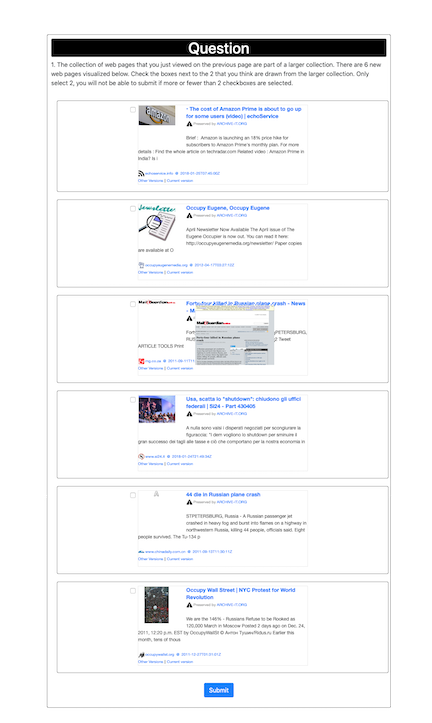}
\caption{Screenshot of question using sc\textasciicircum t surrogates for \emph{Russia Plane Crash Sept 7,2011}}
\end{figure*}

\begin{figure*}[htbp]
\includegraphics[height=8in]{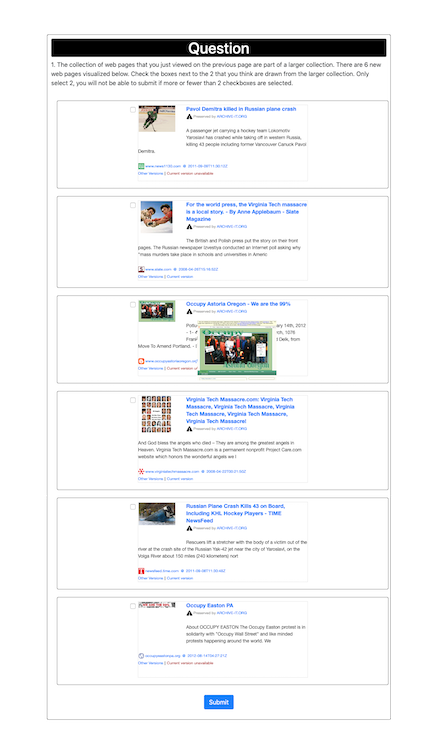}
\caption{Screenshot of question using sc\textasciicircum t surrogates for \emph{Occupy Movement 2011/2012}}
\end{figure*}

\clearpage
\subsection{Completion Code}

\begin{figure*}[htbp!]
\centering
\includegraphics[width=\textwidth]{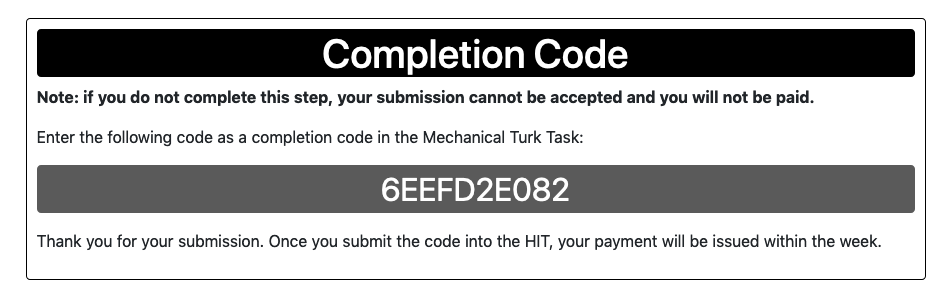}
\caption{Screenshot of a completion code given to study participants. Note that each participant received a different completion code.}
\end{figure*}

\end{document}